\documentclass[pdflatex,sn-mathphys-num]{sn-jnl}

\usepackage{graphicx}%
\usepackage{multirow}%
\usepackage{amsmath,amssymb,amsfonts}%
\usepackage{amsthm}%
\usepackage{mathrsfs}%
\usepackage[title]{appendix}%
\usepackage{xcolor}%
\usepackage{textcomp}%
\usepackage{manyfoot}%
\usepackage{booktabs}%
\usepackage{algorithm}%
\usepackage{algorithmicx}%
\usepackage{algpseudocode}%
\usepackage{listings}%
\usepackage{todonotes}
\usepackage{epstopdf} %
\newcommand{\rev}[1]{\textcolor{black}{#1}}
\newenvironment{revision}{\begingroup\color{black}}{\endgroup}

\theoremstyle{thmstyleone}%
\theoremstyle{thmstyletwo}%

\theoremstyle{thmstylethree}%

\begin{document}

\title[The Fermi-Pasta-Ulam-Tsingou problem after 70 years: toward universal laws of thermalization in lattice systems in the thermodynamic limit]{The Fermi-Pasta-Ulam-Tsingou problem after 70 years: toward universal laws of thermalization in lattice systems in the thermodynamic limit}


\author*[1]{\fnm{Weicheng} \sur{Fu}}\email{fuweicheng@tsnu.edu.cn}

\author[2]{\fnm{Zhen} \sur{Wang}}

\author[3]{\fnm{Wei} \sur{Lin}}
\author[3]{\fnm{Dahai} \sur{He}}
\author[3]{\fnm{Jiao} \sur{Wang}}
\author*[3]{\fnm{Yong} \sur{Zhang}}\email{yzhang75@xmu.edu.cn}
\author*[3]{\fnm{Hong} \sur{Zhao}}\email{zhaoh@xmu.edu.cn}

\affil[1]{\orgdiv{Department of Physics}, \orgname{Tianshui Normal University}, \orgaddress{\city{Tianshui}, \postcode{741000}, \state{Gansu}, \country{China}}}

\affil[2]{\orgdiv{Institute of Theoretical Physics}, \orgname{Chinese Academy of Sciences}, \orgaddress{\city{Beijing}, \postcode{100190}, \country{China}}}

\affil[3]{\orgdiv{Department of Physics}, \orgname{Xiamen University}, \orgaddress{\city{Xiamen}, \postcode{361005}, \state{Fujian}, \country{China}}}


\abstract{
\begin{revision}
The Fermi--Pasta--Ulam--Tsingou (FPUT) problem provides a paradigmatic framework for investigating whether and how weakly nonlinear many-body Hamiltonian systems approach energy equipartition. This focused review briefly outlines the major developments in the FPUT problem, including finite-dimensional near-integrable dynamics, wave resonances, and phonon kinetic theory, and emphasizes recent theoretical analyses, numerical tests, and extensions of scaling laws for thermalization times in nonlinear lattices. Building on accumulated theoretical and numerical evidence, we synthesize a coherent picture of lattice thermalization in the thermodynamic limit: according to the eigenmode properties of an appropriate integrable reference system, the lattices studied so far can be organized into two classes. In the first class, the integrable reference system possesses extended eigenmodes. When the effective perturbation is sufficiently weak and the leading relevant resonant or quasi-resonant processes form a sufficiently connected network in the thermodynamic limit, the thermalization time follows $T_{\mathrm{eq}}\propto g^{-2}$, where $g$ measures the effective deviation from the appropriate integrable reference system. This inverse-square scaling applies broadly to a variety of ordered and weakly disordered lattice models in one, two, and three dimensions, and is robust against changes in the interaction potential, the mechanism that breaks integrability, and the choice of multimode initial states. Correctly identifying the integrable reference point is essential for observing this scaling. This issue is particularly important in one dimension: for certain FPUT-type lattices with cubic interactions, a nearby nonlinear Toda lattice, rather than the harmonic chain, provides the appropriate reference. Otherwise, characterizing the perturbation solely by the bare nonlinear coefficient or the energy density can produce apparent scaling laws that depend on the model and parameter regime. In the second class, all eigenmodes of the integrable reference system are localized. As the perturbation is reduced, spatial-overlap constraints among localized modes progressively fragment low-order resonance networks, and thermalization becomes controlled by higher-order resonant processes. Within the models and finite parameter ranges studied so far, numerical simulations reveal successive scaling regimes of the form $T_{\mathrm{eq}}\propto g^{-\gamma}$, and $\gamma=2,4,6$,
together with a much weaker system-size dependence than in lattices containing extended modes. These results identify a mechanism by which thermalization is progressively suppressed in fully localized systems. Whether this hierarchy of higher-order scaling regimes persists in the asymptotically weak-perturbation limit, and whether a genuinely nonzero thermalization threshold exists, remain open questions. We also review the perturbative wave-resonance framework and resonance-network analysis underlying this picture, and discuss unresolved issues concerning finite-size effects, strongly nonintegrable dynamics, heat transport, localization, and higher-dimensional lattices.
\end{revision}
}


\maketitle

\section{Introduction}\label{sec1}

\begin{revision}
The Fermi--Pasta--Ulam--Tsingou (FPUT) problem arose from a fundamental question concerning the dynamical foundations of statistical mechanics: whether and how a weakly nonintegrable many-body Hamiltonian system approaches equilibrium. In the original numerical experiment, energy initially placed in a small number of low-frequency modes did not spread irreversibly over all normal modes but instead exhibited an unexpected recurrence \cite{Fermi:1955}. This observation initiated the computational study of nonlinear dynamics and stimulated major developments in soliton theory, integrable systems, Hamiltonian chaos, discrete breathers, and anomalous transport \cite{Ford:1992,Campbell:2005,gallavotti_fermi-pasta-ulam_2008}. The continuing interest in the FPUT problem reflects the coexistence of several dynamical regimes whose relevance depends on system size, nonlinear strength, initial conditions, and observation time.

One major line of research treats the FPUT problem as a finite-dimensional, nearly integrable Hamiltonian system. The Kolmogorov--Arnold--Moser (KAM) theorem establishes the persistence of a positive-measure set of invariant tori under sufficiently weak perturbations, subject to suitable nondegeneracy and Diophantine conditions \cite{Kolmogorov1954,arnol1963small,moser1962invariant}. Under appropriate analyticity and steepness conditions, Nekhoroshev theory further provides stretched-exponentially long stability estimates for the actions of nearly integrable systems \cite{Nekhoroshev:1977}. These results provide a natural framework for understanding recurrence and long-lived metastable behavior in finite systems at low energy.

For FPUT chains, normal-form and symmetry analyses have considerably refined this picture. Rink showed that the resonance and symmetry properties of periodic FPUT chains allow the Hamiltonian, in most cases, to be viewed as a perturbation of a nondegenerate Liouville-integrable normal form, thereby implying the existence of many quasiperiodic KAM motions \cite{Rink2001}. He later proved Nishida's conjecture that almost all sufficiently low-energy motions of anharmonic FPUT lattices with fixed endpoints are quasiperiodic \cite{Rink2006}. Henrici and Kappeler derived fourth-order normal forms for odd and even periodic chains and clarified their integrability and KAM properties \cite{HenriciKappeler2008Normal,HenriciKappeler2009Resonant}. Closely related studies have exploited the proximity of FPUT dynamics to the Toda lattice and to KdV-type continuum descriptions. A recent survey by Bambusi, Carati, and Maiocchi reviews rigorous results for finite-particle systems, the limit of large particle number, continuum approximations, and slow relaxation in the thermodynamic limit \cite{BambusiCaratiMaiocchi2026}. Taken together, these developments show why conclusions obtained at fixed finite size cannot automatically be transferred to a large-system kinetic regime: the limits of weak nonlinearity, large system size, and long time need not commute.

This sensitivity to the dynamical regime is also evident in numerical studies of equipartition times. Some studies have reported a stretched-exponential increase of the equipartition time as the energy density decreases, consistent with a Nekhoroshev-type stability scenario \cite{Berchialla:2004}. Other studies have instead yielded power-law behavior with model- and regime-dependent exponents \cite{deluca1995energy,Parisi:1997,Luca:1999,Benettin:2013}. A particularly instructive finite-size analysis found stretched-exponential behavior at fixed $N$ but a crossover to a power law as the thermodynamic limit was approached \cite{Benettin:2011}. Even within the same model, the apparent scaling can change with system size, energy range, initial excitation, and the diagnostic used to define relaxation \cite{Benettin:2011,matsuyama2015multistage}. Rather than representing a single contradiction, these results indicate that the observed thermalization law depends sensitively on the order of limits and on the dynamical regime being probed. They also illustrate why extracting a general thermalization law directly from finite-time numerical data is difficult.

A complementary line of research describes the statistical redistribution of energy among a large number of weakly interacting modes. In this setting, the phonon Boltzmann equation---a wave-kinetic equation for lattice vibrations---provides a mesoscopic description of resonant energy transfer. The connection between weakly anharmonic lattice dynamics and the phonon Boltzmann equation was developed systematically by Spohn and by Lukkarinen and collaborators \cite{Spohn2006Phonon,AokiLukkarinenSpohn2006,Lukkarinen2016}. These studies clarified the structure of collision operators, their conservation laws and entropy production, and their relation to microscopic transport. Direct comparisons between oscillator dynamics and Boltzmann--Peierls kinetics further showed that a kinetic equation can accurately describe a substantial stage of relaxation even when slower post-kinetic processes are not captured by the same equation \cite{MendlLuLukkarinen2016}. A recent review by Germain, La, and Menegaki surveys the microscopic-to-kinetic and kinetic-to-hydrodynamic limits for nonlinear oscillator chains, the current mathematical status of these limits, and the major open problems \cite{GermainLaMenegaki2026}.

A wave-kinetic description is formulated for a joint regime of weak nonlinearity, large volume, and long kinetic times and generally relies on statistical closure assumptions whose precise form depends on the model. Its applicability also depends on the relation between the discrete mode spacing and nonlinear resonance broadening. In sufficiently large systems, resonant and quasi-resonant interactions can form dense networks through which energy spreads among modes. For a fixed finite lattice, by contrast, the resonance spectrum remains discrete, and higher-order exact resonances or near-integrable dynamics can become dominant as the perturbation decreases. The continuum kinetic description emphasized in the phonon Boltzmann literature and the finite-dimensional resonance analysis used for discrete FPUT chains therefore address complementary regimes within the broader wave-turbulence framework.

Within this broader framework, resonant-interaction methods were applied in detail to FPUT thermalization by Onorato, Lvov, and collaborators. For finite $\alpha$-FPUT chains, Onorato \textit{et al.}\ identified connected six-wave resonances as the leading route to equipartition in the weakly nonlinear discrete regime and predicted the associated long relaxation timescale \cite{Onorato:2015}. Lvov and Onorato subsequently studied a finite $\beta$-FPUT chain and found a crossover between a discrete six-wave regime and a frequency-overlap regime with a different relaxation scaling \cite{Lvov:2018}. For the finite FPUT chains considered, these works demonstrated the importance of exact resonances, nonlinear frequency broadening, and the distinction between discrete and overlap regimes. The broader development of the wave-turbulence approach to one-dimensional chains has recently been reviewed by Onorato, Lvov, Dematteis, and Chibbaro \cite{Onorato:2023}.

Building on wave-resonance theory and its earlier applications to finite FPUT chains, we extended the analysis to the thermodynamic limit and proposed a universal scaling of the thermalization time for a broad class of one-dimensional lattices \cite{Fu:2019R}. The analysis showed that, when the effective perturbation strength $g$ is defined consistently, the leading nonvanishing resonant contribution gives
\begin{equation}
  T_{\mathrm{eq}} \propto g^{-2}.
  \label{eq:intro_inverse_square}
\end{equation}
The scaling was derived for general power-law interactions, and a unified expression was proposed for general one-dimensional interaction potentials. The prediction was tested through large-scale simulations and finite-size analysis. A related study by Pistone \textit{et al.}\ subsequently discussed common features of thermalization in the $\alpha$-FPUT, $\beta$-FPUT, and discrete nonlinear Klein--Gordon chains in both discrete and thermodynamic regimes \cite{Pistone:2019}. Together, these developments connected the resonant-interaction picture for individual models to the search for general thermalization laws.

A key requirement for observing the inverse-square law was to identify the appropriate integrable reference system relative to which the perturbation strength should be measured. For one-dimensional lattices containing a cubic interaction, the nonlinear Toda lattice can provide a more appropriate reference point than the harmonic chain. In the models studied, defining the nonintegrability strength by the deviation from the Toda reference revealed the same inverse-square law in cases that otherwise appeared to exhibit different energy-density exponents \cite{Fu:2019}. This reference-point formulation provides a unified way to compare thermalization in different near-integrable lattices while leaving room for additional effects associated with finite size, initial conditions, or different resonance structures.

The resulting framework has been examined in increasingly broad settings. Studies of perturbed Toda and diatomic lattices showed that different mechanisms of breaking integrability can produce different microscopic relaxation pathways while preserving the same scaling with the effective perturbation strength \cite{Fu:2019,PhysRevE.100.052102}. Studies of mass-disordered chains found that weak disorder can relax wave-vector selection constraints and, in the specific models and parameter regimes considered, accelerate resonant thermalization \cite{Wang:2020,Sun:2020}. Simulations of two- and three-dimensional lattices found the same inverse-square scaling across several lattice structures and interaction potentials while revealing additional roles for mode degeneracy and dimensionality in the energy-transfer process \cite{Wang:2024}.

These extensions also brought the normal-mode structure of the chosen integrable reference system into the description of thermalization. In lattices containing extended modes, the number of accessible resonant and quasi-resonant couplings can grow with system size, favoring a connected resonance network and the inverse-square scaling in the large-system limit. When all reference modes are localized, spatial-overlap restrictions can instead make the connectivity of low-order resonance networks essentially size independent and cause those networks to fragment as the perturbation is reduced. Numerical simulations then reveal successive higher-order relaxation regimes of the form
\begin{equation}
  T_{\mathrm{eq}} \propto g^{-\gamma},
  \qquad \gamma=2,4,6,\ldots,
  \label{eq:intro_thermalization_classes}
\end{equation}
where the listed exponents represent successive perturbative regimes observed over finite parameter ranges rather than a single exponent valid for all $g$ \cite{Lin:2025}. These observations motivated a proposed organization of lattice systems into two thermalization classes based on resonance-network connectivity and on whether the reference-mode spectrum contains extended states. This classification is a theoretically motivated framework supported by resonance-network analysis and numerical evidence, not an exhaustive theorem for all classical lattices. Whether the sequence of higher-order regimes persists asymptotically as $g\to0$, or whether a genuine nonzero thermalization threshold exists in fully localized systems, remains an open question.
Flach and collaborators previously proposed an insightful general classification based on long-range and short-range coupling networks in action space, providing a unified perspective on distinct regimes of chaotic spreading and relaxation in weakly nonintegrable many-body systems~\cite{Danieli:2019,MalishavaFlach:2022,ZhangLandoDietzFlach:2024}. The two classes of thermalizing systems proposed here are consistent with this framework. More specifically, by using the presence or absence of extended eigenmodes in the spectrum of the integrable reference system as the criterion, we provide a clear and operational distinction between the two thermalization behaviors in lattice systems.

Several reviews have surveyed the FPUT problem from historical, dynamical-systems, wave-turbulence, and kinetic-theory perspectives \cite{gallavotti_fermi-pasta-ulam_2008,Onorato:2023,BambusiCaratiMaiocchi2026,GermainLaMenegaki2026}. Against this broader background, the present article is intended as a focused review of the development of scaling laws for thermalization times in nonlinear lattices, beginning with our proposal and covering their wave-resonance basis, subsequent tests and extensions, and connections to related developments. Section~2 introduces the perturbative framework and the roles of resonant and quasi-resonant processes. Section~3 reviews the inverse-square law and its verification in ordered, disordered, one-dimensional, and higher-dimensional systems. Section~4 discusses resonance-network connectivity and the proposed thermalization classes. Section~5 summarizes the main conclusions and outlines open questions concerning finite systems, strongly nonlinear dynamics, transport, localization, and higher-dimensional lattices.
\end{revision}

\section{Perturbation theory analysis}\label{sec2}

\begin{revision}
\subsection{General weak-perturbation formulation}

The Hamiltonian of a nonlinear lattice system can generally be written as
\begin{equation}\label{eq:H0h}
H = H_0 + gV,
\end{equation}
where $H_0$ is Liouville integrable, $V$ is a specified regular perturbation, and $g$ is a small dimensionless control parameter.

Let $\{I_k\}$ be a complete set of actions of $H_0$, so that $\{H_0,I_k\}=0$ with $\{\cdot,\cdot\}$ denoting the Poisson bracket.
In the presence of the perturbation, the dynamics of the integrals of motion follow the Liouville equation \cite{Zwanzig2001}
\begin{equation}\label{eq:ILE}
\partial_t I_k =(L_0+L')I_k= L_0 I_k + L' I_k,
\end{equation}
where $L_0$ and $L'$ are Liouville operators defined by
\begin{equation}
L_0 I_k = \{H_0,I_k\}, \qquad
L' I_k = g \{V,I_k\}.
\end{equation}
A standard second-order projection or cumulant expansion gives, after a separation between the fast $H_0$ dynamics and the slow evolution of the actions,
\begin{align}\label{eq:kinetic_equation-1}
\partial_t\langle I_k\rangle
={}&\langle L'I_k\rangle_{f(0)}
+\int_0^{\infty}\!\langle L'L'(\tau)I_k\rangle_{f(t)}\,d\tau
+O(g^3),
\end{align}
where
\begin{equation}
L'(\tau)=e^{-L_0\tau}L'e^{L_0\tau},
\end{equation}
and $f(t)$ denotes the phase-space distribution of the system described by Hamiltonian \eqref{eq:H0h}. The notation $\langle I_k\rangle$ represents the ensemble average with respect to $f(t)$.
For sufficiently weak perturbations, the dynamics are dominated by $H_0$, and the slowly evolving quantities $I_k(t)$ may be regarded as adiabatic deformations of the integrals of motion of $H_0$. In this regime the distribution can be approximated by a generalized Gibbs ensemble \cite{Wang_2024CTP},
\begin{equation}\label{ff}
f(t) \sim \exp\!\left[-\sum_k \theta_k(t) I_k\right],
\end{equation}
where $\theta_k$ are the Lagrange multipliers associated with $I_k$ \cite{Dudnikova2003}. The perturbation therefore drives the slow evolution of the system within the space of the conserved quantities of $H_0$.

In the phase-randomized situations considered here, the first-order secular contribution vanishes. The second-order term scales as
$g^2$ provided that the relevant correlation integral converges, the decorrelation or random-phase approximation holds, and the perturbation couples the actions via connected processes. These constitute physical assumptions rather than direct implications of Liouville integrability alone. Under the validity of perturbation theory, the kinetic timescale for relaxation of the actions scales inversely with $g^2$, as follows from the kinetic equation for the actions. For a concrete Hamiltonian system, detailed analysis within the perturbative framework is required to examine its applicability.

Note that the kinetic equation is not asserted to be an exact equation for an arbitrary finite lattice. It is an asymptotic mesoscopic description obtained through a singular combination of weak coupling, large system size, and long time. Schematically, one studies $g\to0$ while resolving the kinetic time $t=g^{-2}\tau$, and simultaneously takes a large-box or thermodynamic limit so that the relevant frequency sums approach resonant integrals. The derivation additionally uses statistical assumptions such as random phases, quasi-Gaussian closure or propagation of chaos, and sufficient mixing of the reference dynamics \cite{Spohn2006Phonon,GermainLaMenegaki2026}.
These limits need not commute. At fixed $N$, the frequency spectrum is discrete, exact resonances may be absent or form disconnected clusters, and near-resonant terms that are negligible in one scaling regime may dominate in another. Conversely, taking $N\to\infty$ first can create a dense resonant manifold that is not present in a finite chain. Recent work on the $\beta$-FPUT normal-form transformation makes this issue explicit by finding a size-dependent near-identity condition of the form $\beta\ll N^{-1-\delta}$ for random-phase fields \cite{WuOnoratoHaniPan2025}. Rigorous derivations for FPUT currently cover only restricted times or reduced dynamics, unlike the more complete results obtained by Deng and Hani for NLS \cite{DengHani2021,DengHani2023,Wu2025WKE}. We therefore use the kinetic equation below as a conditional asymptotic theory and compare its predictions with finite-$N$ simulations only after checking size convergence and resonance connectivity.

\subsection{Explicit normal-mode specialization for a quadratic reference}

For the explicit homogeneous-chain examples used below, let $r_j = x_{j+1}-x_j$ denote the relative displacement between adjacent particles, where $x_j$ represents the displacement of the $j$-th particle from its equilibrium position. When the reference integrable system is the harmonic chain,
\begin{equation}
H_0^{\rm har}=\sum_{j=1}^{N}\left(\frac{p_j^2}{2}+\frac{r_j^2}{2}\right),
\end{equation}
where $p_j$ is the momentum of the $j$th particle, $N$ is the total number of particles, and one analytic monomial perturbation is
\begin{equation}\label{eq:Hnprime}
H_n'=\frac{\lambda_n}{n}\sum_{j=1}^{N} r_j^n,
\qquad n=3,4,\ldots .
\end{equation}
For odd $n$, Eq.~\eqref{eq:Hnprime} is understood as a term in a full stable interaction potential rather than as a globally stable potential by itself. A separate numerical comparison in Ref.~\cite{Fu:2019R} used the symmetric family
\begin{equation}\label{eq:Hdsym}
H_{d,\rm sym}'=\frac{\lambda_d}{d}\sum_{j=1}^{N} |r_j|^d,
\qquad d>2.
\end{equation}
The exponent $d$ is not a typographical substitute for $n$: Eqs.~\eqref{eq:Hnprime} and \eqref{eq:Hdsym} denote different model families. The absolute-power family is even under $r_j\mapsto-r_j$ and, for non-even $d$, is not analytic at the origin to all orders. We therefore use it only as a deliberately symmetric comparison model, not as the Taylor expansion of a generic molecular potential and not in the analytic monomial derivation.

For a homogeneous term of degree $n$ around the harmonic reference, the energy rescaling $r_j=\sqrt{\varepsilon}\,\tilde r_j$ gives
\begin{equation}\label{eq:gn}
g=\lambda_n\varepsilon^{(n-2)/2},
\end{equation}
and analogously $g=\lambda_d\varepsilon^{(d-2)/2}$ for Eq.~\eqref{eq:Hdsym}, where $\varepsilon$ is the energy density (i.e., energy per particle). Equation~\eqref{eq:gn} is thus a specialization to a quadratic $H_0$ and a homogeneous perturbation; it is not a general definition for an arbitrary integrable reference Hamiltonian.

For lattice chains, it is convenient to introduce canonical complex normal variables
\begin{equation}
a_k = \frac{P_k - i\omega_k Q_k}{\sqrt{2\omega_k}},
\end{equation}
where $Q_k$ and $P_k$ are the Fourier transforms of the displacements and momenta, respectively, and $\omega_k$ denotes the dispersion relation associated with wave number $k$. Note that $k$ has a direct physical meaning only in translationally invariant systems; otherwise it merely serves as a label ordering the modes.
The energy of the $k$th mode is
\begin{equation}\label{eq:Ek}
E_k = \omega_k a_k a_k^* = \omega_k I_k .
\end{equation}
Accordingly, the integrable Hamiltonian can be written as
\begin{equation}\label{eq:H0}
H_0^{\rm har}= \sum_k E_k = \sum_k \omega_k I_k .
\end{equation}
Only when $H_0$ is quadratic can its actions be identified directly with independent harmonic eigenmodes. When $g=0$, Eq.~\eqref{eq:ILE} implies that each integral of motion $I_k$ is conserved. Consequently, the modal energies $E_k$ remain constant and are completely determined by the initial state. For nonzero $g$, the perturbation can couple these integrals of motion. When the relevant resonant or quasi-resonant processes are nonvanishing and form a sufficiently connected network, the system may approach a thermalized state. We define the time-averaged modal energy as
\begin{equation}\label{eq:aEk}
\bar{E}_k(t)=
\frac{1}{t}
\int_0^{t} E_k(t')\, dt',
\end{equation}
and at equipartition $\bar{E}_k(T_{\mathrm{eq}})\simeq\varepsilon$ for all modes, where $T_{\mathrm{eq}}$ denotes the thermalization time required for the system to evolve from its initial state to the equipartition state.

Based on the defined $\bar{E}_k(t)$, one can introduce a parameter to measure how close the system is to equipartition. A frequently used parameter is the effective relative number of degrees of freedom $\xi(t)$ as in Ref.~\cite{Benettin:2011}, i.e.,
\begin{equation}
\xi(t) = \tilde{\xi}(t)\frac{\mathrm{e}^{\eta(t)}}{N/2},
\label{eq:xi_def}
\end{equation}
where
\begin{equation}
\eta(t) = -\sum_{k=N/2}^{N} w_k(t)\log\big[w_k(t)\big]
\label{eq:spectral_entropy}
\end{equation}
is the spectral entropy and
\begin{equation}
\tilde{\xi}(t) = \frac{\sum_{k=N/2}^{N}\bar{E}_k(t)}{\frac{1}{2}\sum_{1\leqslant k\leqslant N}\bar{E}_k(t)},\quad
w_k(t) = \frac{\bar{E}_k(t)}{\sum_{j=N/2}^{N}\bar{E}_j(t)}.
\label{eq:tilde_xi_wk}
\end{equation}
When equipartition is approached, $\xi$ will saturate at 1.

Under the kinetic closure, the mode action takes the relaxation form
\begin{equation}\label{eq:kenetic_equation}
\partial_t\langle I_k\rangle=\eta_k-\gamma_k\langle I_k\rangle,
\end{equation}
with $\eta_k,\gamma_k=O(g^2)$ \cite{Wang_2024CTP}. If $\gamma_k \neq 0$, Eq.~\eqref{eq:kenetic_equation} implies that $\langle I_k\rangle$ relaxes toward equilibrium on a modal timescale
\begin{equation}
\tau_k \sim \gamma_k^{-1} \propto g^{-2}.
\end{equation}
A global thermalization time $T_{\mathrm{eq}}\propto g^{-2}$ follows only when the kinetic closure remains valid and these modal processes form a sufficiently connected network for global energy redistribution.

\end{revision}

\begin{revision}
For systems with translational symmetry, a nonvanishing collision term requires resonant processes satisfying the $n$-wave frequency and lattice-momentum conditions
\begin{align}
\omega_1+\dots+\omega_\ell
&=
\omega_{\ell+1}+\dots+\omega_n,
\label{eq:MWRC_w}
\\
k_1+\dots+k_\ell
&=
k_{\ell+1}+\dots+k_n+G ,
\label{eq:MWRC_k}
\end{align}
where $G$ is a reciprocal-lattice vector and includes Umklapp processes. These resonance conditions alone do not guarantee $\gamma_k\ne0$: the corresponding interaction matrix elements and collision integral must also be nonvanishing, and the resonant processes must provide sufficient connectivity across mode space. In large systems the dispersion frequencies $\omega_k$ become densely distributed. Moreover, nonlinear interactions lead to a finite frequency broadening $\Omega$ \cite{2011LNP825N}. At finite broadening, the exact frequency condition \eqref{eq:MWRC_w} is therefore supplemented by the quasi-resonant criterion
\begin{equation}\label{eq:quasi-MWRC_w}
\left|
\omega_1+\dots+\omega_\ell
-
\omega_{\ell+1}-\dots-\omega_n
\right|
\lesssim \Omega.
\end{equation}

The same perturbative approach can in principle be extended to higher-dimensional systems such as two- and three-dimensional lattices, although the resonance structure, spectral degeneracies, and connectivity conditions become dimension dependent \cite{Wang:2024}.
\end{revision}

\begin{revision}
\subsection{Nonlinear integrable reference systems: the perturbed Toda chains}\label{sec2-subToda}

For a nonlinear integrable reference such as the Toda chain, action--angle variables exist for finite systems \cite{PhysRevB.82.104306}, but the simple representation $H_0=\sum_k\omega_kI_k$ with amplitude-independent frequencies and the explicit harmonic-mode resonance network above do not apply. The general second-order expression \eqref{eq:kinetic_equation-1} suggests an $O(g^{-2})$ kinetic timescale if its correlation and connectivity assumptions can be established, but the present review does not derive a closed Toda-action kinetic equation or a connected resonance network from first principles. The perturbed-Toda results summarized in Sec.~\ref{sec3} should therefore be read as perturbative reference-system analysis supported by numerical scaling, not as a direct corollary of Eqs.~\eqref{eq:Ek}--\eqref{eq:quasi-MWRC_w}.

The choice of the closest Toda reference is especially important for a cubic perturbation. With
\begin{equation}
V_{\rm T}(\alpha,r)=\frac{e^{2\alpha r}-2\alpha r-1}{4\alpha^2}
=\frac{r^2}{2}+\sum_{m=3}^{\infty}\frac{c_m(\alpha)}{m}r^m,
\qquad
c_m(\alpha)=\frac{(2\alpha)^{m-2}}{(m-1)!},
\end{equation}
one obtains
\begin{equation}\label{eq:toda-shift}
V_{\rm T}(\alpha,r)+\frac{\theta_3}{3}r^3
=V_{\rm T}(\tilde\alpha,r)
+\sum_{m=4}^{\infty}\frac{c_m(\alpha)-c_m(\tilde\alpha)}{m}r^m,
\qquad \tilde\alpha=\alpha+\theta_3.
\end{equation}
Thus the residual perturbation begins at fourth order, with coefficient
$[\alpha^2-\tilde\alpha^2]/6$ multiplying $r^4$. For the numerical plots we denote the bare added coefficient by $\theta_n$ and the associated dimensionless residual strength by $\epsilon_n$, see Fig.~\ref{fig19NJP}. For $n\ge4$,
$\epsilon_n=|\theta_n|\varepsilon^{(n-2)/2}$;
for the cubic case, $\epsilon_m=|c_m(\alpha)-c_m(\tilde\alpha)|\varepsilon^{(m-2)/2}$ and the leading quantity is $\epsilon_4$ (i.e., $m=4$). This is the construction presented in Ref.~\cite{Fu:2019}.

\section{Numerical evidence for inverse-square thermalization laws}\label{sec3}

The inverse-square relation should be understood as a kinetic-regime statement with explicit conditions. We write $H=H_0+gV$ only after specifying the reference integrable system and the normalization of $V$. When the second-order correlation integral is finite and the lowest relevant quasi-resonance network is connected, the relaxation rate is $O(g^2)$ and $T_{\mathrm{eq}}\propto g^{-2}$. A crucial requirement for the validity of this relation is that the integrable reference Hamiltonian $H_0$ must be chosen appropriately. The choice of $H_0$ determines both the explicit form of perturbation and magnitude of the effective perturbation strength $g$, and therefore directly affects the quantitative characterization of the system's thermalization capability. Finite-size frequency discreteness, special initial packets, symmetry constraints, incomplete network connectivity, or a nonoptimal reference Hamiltonian can instead produce crossovers or different apparent exponents.

\begin{figure}[t]
\centering
\includegraphics[width=1\columnwidth]{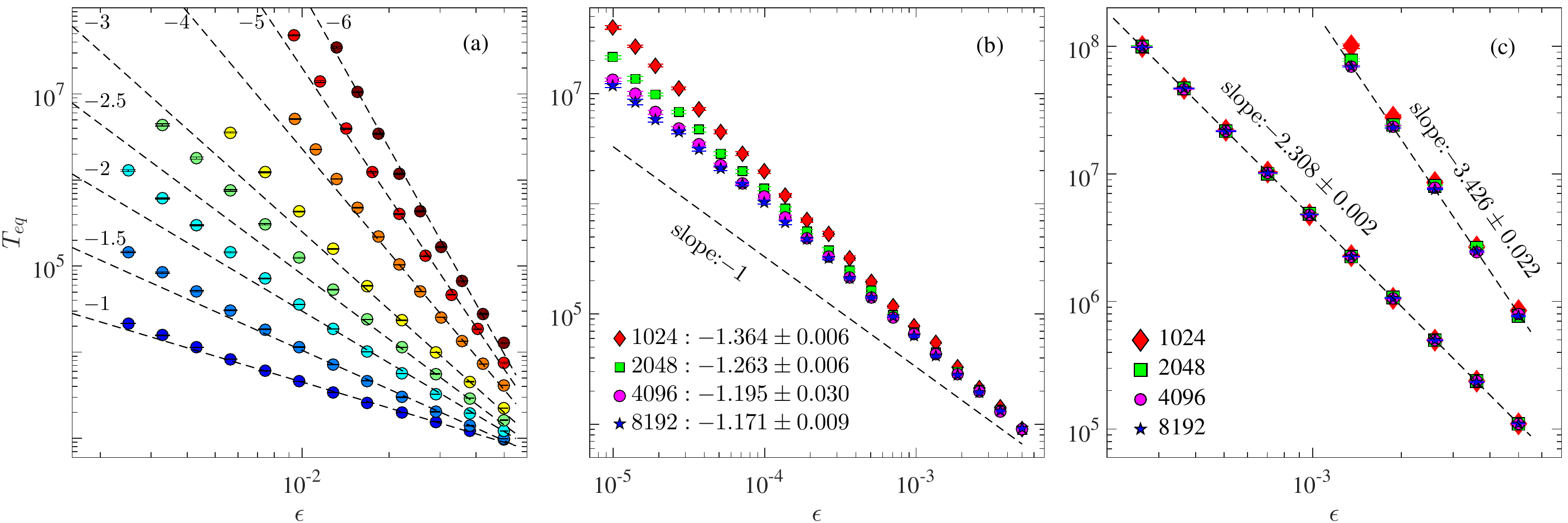}
\caption{\rev{Thermalization time versus energy density for power-law perturbations of the harmonic chain with fixed boundary conditions. To suppress fluctuations, the average is done over 60 phases uniformly distributed in $[0, 2\pi]$. (a) Symmetric comparison models $V(r)=r^2/2+|r|^d/d$, with $d=3,7/2,4,9/2,5,6,7,8$ from bottom to top and $N=2048$; the dashed lines have the theoretical reference slopes $2-d$ and are not fitted lines. (b) Finite-size convergence for the symmetric $d=3$ model, with $N=1024,2048,4096,8192$; the numerical least-squares slopes are printed in the panel. (c) Analytic asymmetric models $V(r)=r^2/2+r^n/n$ for $n=3$ and $5$; the dashed lines are least-squares fits to the $N=8192$ data, with slopes $-2.308\pm0.002$ and $-3.426\pm0.022$, respectively. The symbols and exponents $d$ and $n$ therefore refer to two distinct model families. The error bars are much smaller than the symbols. Reprinted from Ref.~\cite{Fu:2019R}.}
}\label{Fig2019R}
\end{figure}

\end{revision}

The various power-law exponents for thermalization reported in early studies can largely be attributed to two factors. The first is numerical bias arising from finite system sizes and limited computational resources. The second is the improper choice of the reference integrable Hamiltonian, which leads to a distorted definition of the effective perturbation strength.
Extensive studies have shown that once an appropriate integrable reference system $H_0$ is chosen and the perturbation strength $g$ is consistently defined, the scaling relation $T_{\mathrm{eq}}\propto g^{-2}$ exhibits remarkable robustness \cite{Fu:2019R,Fu:2019,PhysRevE.100.052102,Wang:2020,Fu:2021,Feng_2022}.
Importantly, this conclusion is largely insensitive to the specific mechanism by which integrability is broken. For example, integrability may be destroyed by introducing nonlinear interactions \cite{Fu:2019,Fu:2019R}, by incorporating mass disorder \cite{PhysRevE.100.052102,Wang:2020}, or through stress/strain modulation \cite{Fu:2021}.
Moreover, the same scaling behavior is observed regardless of whether the initial excitation is placed in low-frequency modes or high-frequency modes (the so-called anti-FPUT scenario) \cite{Feng_2022}. In all these cases, the thermalization time follows the same inverse-square scaling once the perturbation strength is defined within a consistent perturbation framework.

\begin{revision}

The normal-mode equations under different boundary conditions retain the same generic multi-wave-coupling structure \cite{BIVINS197365,SHOLL1990253}. The boundary conditions nevertheless modify the allowed wave numbers, the sampling and degeneracy of the linear spectrum, and the associated resonance-selection rules and effective scattering phase space. In particular, periodic boundary conditions introduce the degeneracy $\omega_k=\omega_{-k}$. For the same number of particles, a periodic chain therefore contains fewer distinct mode frequencies than its fixed-boundary counterpart. Translational invariance under periodic boundary conditions also imposes wave-number conservation on the scattering processes, reducing the number of independent scattering channels available at a fixed system size. These spectral and kinematic features generally produce stronger finite-size effects under periodic boundary conditions. Accordingly, to obtain a denser sampling of the linear spectrum at a fixed number of particles, and thereby reduce finite-size restrictions on the quasi-resonance network, we employ fixed boundary conditions in our molecular-dynamics simulations.

All numerical results presented in this work were obtained under fixed boundary conditions, with the lowest-frequency $10\%$ of the normal modes initially excited. We have also verified that, for multimode initial states, varying the fraction of initially excited modes or exciting a high-frequency rather than a low-frequency band does not alter the observed scaling behavior \cite{Feng_2022}. We employ the effective relative number of degrees of freedom $\xi(t)$ as an indicator to characterize the scaling behavior of $T_{\mathrm{eq}}$ with respect to the perturbation strength $g$.
For mass-disordered systems, each disorder realization is generated by independently drawing the particle masses from the uniform distribution
$m_j\in[1-\delta m,1+\delta m]$, where $\delta m$ denotes the disorder strength. For each mass configuration, we numerically diagonalize the corresponding harmonic Hamiltonian to obtain its eigenfrequencies and eigenvectors, which are then used to define the normal modes and determine the associated thermalization time. For a fixed mass configuration, the reported value of $T_{\mathrm{eq}}$ is obtained by averaging over multiple independent choices of the random initial phases. We have verified that, for systems containing on the order of $10^3$ particles or more, the thermalization times obtained from different independently generated mass configurations are nearly indistinguishable, indicating an effective self-averaging behavior. For smaller systems containing only several tens of particles, sample-to-sample fluctuations are more pronounced; therefore, in addition to averaging over random initial phases for each fixed mass configuration, we further perform an ensemble average over multiple independent disorder realizations.

We first consider the simplest setting in which the harmonic Hamiltonian is chosen as the integrable reference system and nonlinear terms of different orders are treated as perturbations \cite{Fu:2019R}. The corresponding thermalization behavior is summarized in Fig.~\ref{Fig2019R}, where the perturbing parts of the interaction potentials are given by Eqs.~\eqref{eq:Hnprime} and \eqref{eq:Hdsym} with fixed $\lambda_n=\lambda_d=1$. As shown in Fig.~\ref{Fig2019R}(a), the thermalization time approximately follows
\begin{equation}
    T_{\mathrm{eq}}\propto\varepsilon^{-(d-2)}\sim g^{-2}.
\end{equation}
For several values of $d$, the numerical data gradually deviate from this prediction as the energy density is decreased. These deviations, however, originate from finite-size effects rather than from a breakdown of the asymptotic scaling law. This point is illustrated particularly clearly by the case $d=3$. In Fig.~\ref{Fig2019R}(a), almost all data points lie close to the predicted line
$T_{\mathrm{eq}}\propto\varepsilon^{-1}$.
Figure~\ref{Fig2019R}(b), however, shows that, at any fixed system size, the thermalization time eventually departs from the $\varepsilon^{-1}$ scaling as $\varepsilon$ is reduced further. Increasing the system size systematically shifts this departure toward lower energy densities and brings the numerical results back toward the theoretical prediction. The observed behavior therefore directly reflects the noncommutativity of the weak-perturbation and thermodynamic limits.
More precisely, for a given finite system, the kinetic scaling
$T_{\mathrm{eq}}\propto\varepsilon^{-(d-2)}\sim g^{-2}$
may be observed only over a finite range of energy densities. As the energy density, or equivalently the effective perturbation strength, is reduced, increasingly larger systems are required to recover the same scaling regime. This is because the kinetic description is an asymptotic large-system theory: the thermodynamic limit must be approached sufficiently rapidly relative to the weak-perturbation limit so that the mode spectrum becomes dense enough and the relevant quasi-resonance network remains effectively connected. Only in this regime is the relaxation dynamics governed by the kinetic equation associated with the perturbation, yielding the characteristic timescale
$T_{\mathrm{eq}}\propto g^{-2}$ \cite{Fu:2019R}.

As shown in Fig.~\ref{Fig2019R}(c), the fitted slopes $-2.308\pm0.002$ and $-3.426\pm0.022$ for the asymmetric models with $n=3$ and $n=5$ deviate from the theoretical values of $-2$ and $-3$, respectively. Although the $n=5$ data exhibit a recognizable system-size dependence, the fitted slope approaches the theoretical prediction only slowly as the system size is increased. This tendency is even more pronounced for $n=3$, for which the size dependence is considerably weaker. In fact, when $n=3$, the system corresponds to the FPUT-$\alpha$ model, which can be regarded as the FPUT-$\alpha$-$\beta$ model with $\beta=0$. The fitted value is very close to the result for the FPUT-$\alpha$-$\beta$ model reported in reference \cite{Benettin:2011}, namely $-9/4 = -2.25$. We attribute this weak finite-size dependence to the asymmetry of the interaction potential. Our phonon-spectrum analysis shows that the spectral peaks of the asymmetric models have asymmetric line shapes and deviate appreciably from the symmetric Lorentzian envelopes observed in the symmetric models \cite{Fu:2019R}. These asymmetric spectral peaks also possess larger linewidths, which relax the frequency-matching condition and make quasi-resonant scattering effective already at relatively small system sizes. Consequently, further increasing the system size has only a limited influence on the quasi-resonant processes, resulting in the weak size dependence observed for the asymmetric models. However, the situation is different for the symmetric models, whose phonon spectral peaks are smooth, symmetric, and well described by Lorentzian envelopes with smaller linewidths. At small system sizes, the discrete frequency spacing then strongly restricts quasi-resonant scattering. Increasing the system size produces a denser mode spectrum and substantially enhances the contribution of quasi-resonant processes, so that the fitted slopes exhibit a clearer tendency to approach the theoretical predictions. We suggest that the remaining deviations of the asymmetric models from the predicted exponents may originate from higher-order resonant processes that remain relevant over the numerically accessible parameter range \cite{Benettin:2011,e27070741,Fu:2019}. A direct decomposition of the energy-transfer rate into different resonance orders would be required to verify this interpretation quantitatively in future work.
\end{revision}

\begin{figure}[t]
\centering
\includegraphics[width=.7\columnwidth]{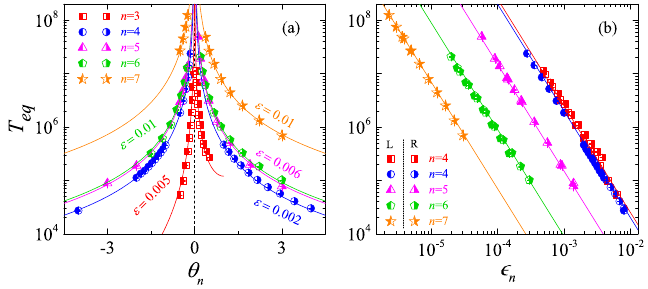}
    \caption{\rev{Perturbed-Toda results with fixed boundary conditions and $\alpha=-1$, $N=2048$. To suppress fluctuations, the average is done over 24 phases
uniformly distributed in $[0, 2\pi]$. Initially the lowest $10\%$ of frequency modes are excited. The interaction is $V_{\rm T}(\alpha,r)+\theta_n r^n/n$, where $\theta_n$ is the bare added coefficient. (a) $T_{\mathrm{eq}}$ versus $\theta_n$ for $n=3,4,5,6,7$; the solid $\Lambda$-shaped curves use the best-fitting prefactors of the residual-strength prediction. (b) The same data versus the dimensionless residual strength. For $n\ge4$, it is $\epsilon_n=|\theta_n|\varepsilon^{(n-2)/2}$. For $n=3$, the reference is first shifted to $V_{\rm T}(\alpha+\theta_3,r)$ and the residual starts at fourth order as in Eq.~\eqref{eq:toda-shift}; the leading plotted residual strength is therefore $\epsilon_4=|c_4(\alpha)-c_4(\alpha+\theta_3)|\varepsilon$. The straight lines of slope $-2$ are reference lines, and L/R denote data on the left/right sides of the peak in panel (a). The error bars are much smaller than the symbols. Reprinted from Ref.~\cite{Fu:2019}.}}\label{fig19NJP}
\end{figure}

\begin{figure}[t]
\centering
\includegraphics[width=.7\columnwidth]{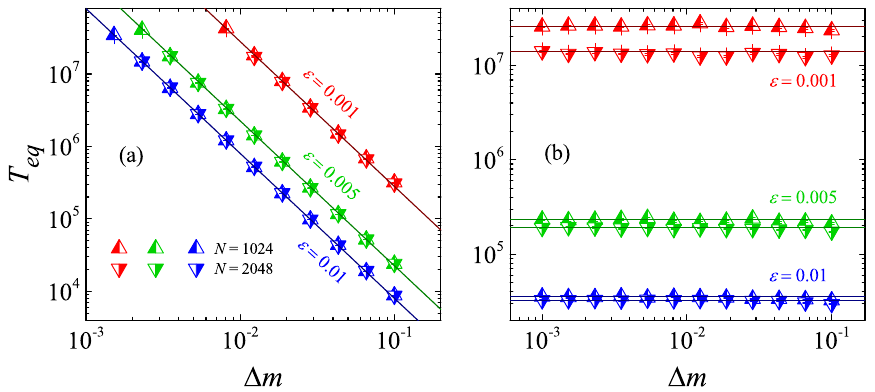}
\caption{Thermalization time in diatomic chains with fixed boundary conditions. To suppress fluctuations, the ensemble average is done over 24 phases uniformly distributed in $[0, 2\pi]$. The lowest $10\%$ of the frequency modes are excited.
(a) Integrability of the nonlinear Toda Hamiltonian is broken by introducing unequal masses, leading to a perturbation whose strength is characterized by the mass difference. The thermalization time follows the inverse-square scaling with the perturbation strength.
(b) Results for the diatomic FPUT-$\beta$ chain shown for comparison. In this case the relevant integrable reference system is the harmonic Hamiltonian with the same mass configuration, and therefore the mass difference does not change the effective perturbation strength. The error bars are much smaller than the symbols. Reprinted from Ref. \cite{PhysRevE.100.052102}.}\label{Fig2019PRE}
\end{figure}

\begin{revision}
Across the broader set of near-integrable examples, inverse-square behavior is observed when the effective perturbation is defined relative to an appropriate integrable reference and the kinetic conditions are reached \cite{Fu:2019R,Fu:2019,PhysRevE.100.052102,Wang:2020,Fu:2021,Feng_2022}.
Figure~\ref{fig19NJP} shows the dependence of the thermalization time on perturbations of different nonlinear orders to the Toda lattice, with $\alpha=-1$ and the system size fixed at $N=2048$. Figure~\ref{fig19NJP}(a) presents the thermalization time as a function of the bare nonlinear coefficient $\theta_n$. The resulting curves exhibit a characteristic $\Lambda$-shaped dependence. By contrast, Fig.~\ref{fig19NJP}(b) plots the same thermalization times against the corresponding effective perturbation strength defined relative to the nearest integrable Toda lattice; see again Subsection~\ref{sec2-subToda} or Ref.~\cite{Fu:2019} for more details. Nearly all data points collapse onto the reference line with slope $-2$, consistent with the scaling $T_{\mathrm{eq}}\sim g^{-2}$ over the parameter range studied.

As shown by the preceding theoretical analysis, adding a cubic term shifts the appropriate integrable reference system from the original Toda lattice with parameter $\alpha$ to another Toda lattice with a renormalized parameter. After this shift is absorbed into the new reference Hamiltonian, the residual nonintegrable perturbation begins at quartic order rather than cubic order. Consequently, when expressed in terms of the effective perturbation strength, the thermalization data for the $\theta_3$ perturbation almost coincide with those obtained for the explicitly quartic $\theta_4$ perturbation. This agreement demonstrates that the thermalization time is controlled not by the nominal order or bare coefficient of the added term, but by the strength of the residual perturbation measured relative to the nearest integrable reference system. In particular, the nominal order of an added interaction term does not necessarily coincide with the leading order of the effective nonintegrable perturbation.

A further clarification is required for the perturbed Toda lattice. In the numerical calculations reviewed here, the thermalization time is extracted from the relaxation of mode energies defined in the harmonic normal-mode basis. In principle, a description adapted to the integrable reference system should instead follow the redistribution of energy among the Toda action variables. This is technically difficult because the Toda actions are represented by nontrivial infinite series~\cite{PhysRevB.82.104306}, and an explicit inverse transformation suitable for preparing controlled real-space initial conditions is not readily available. Nevertheless, previous work directly monitored the relaxation of the first few Toda actions in a perturbed Toda chain and found that their characteristic relaxation times also scale inversely with the square of the perturbation strength~\cite{PhysRevB.82.104306}. This result does not constitute a closed kinetic theory for the full set of Toda actions, but it provides independent numerical support for the inverse-square scaling observed in the relaxation of harmonic-mode energies.

Figure~\ref{Fig2019PRE} shows the thermalization behavior of one-dimensional diatomic Toda and FPUT-$\beta$ chains under fixed boundary conditions, with the lowest-frequency $10\%$ of the normal modes initially excited. In the Toda chain, integrability is broken by the mass inhomogeneity, and the corresponding perturbation strength is characterized by the mass difference between the two alternating species. The thermalization time again follows an inverse-square dependence on this effective perturbation strength. For comparison, panel~(b) presents the corresponding results for the diatomic FPUT-$\beta$ chain. In this intrinsically nonintegrable system, changing the masses does not modify the integrability structure relevant to the perturbative description, because the appropriate integrable reference system is the harmonic Hamiltonian with the same mass configuration. Consequently, the thermalization time remains essentially unchanged as the mass difference is varied. Instead, its relaxation time depends sensitively on the energy density, which determines the effective nonlinear perturbation strength~\cite{PhysRevE.100.052102}.

We have further verified that these conclusions are insensitive to the specific multimode initial condition. Changing the fraction of initially excited modes, or replacing the initially excited low-frequency band by a high-frequency band, can substantially modify the detailed pathway toward equipartition. Nevertheless, the principal conclusions of both panels remain unchanged: the perturbed Toda chain retains the inverse-square scaling with the effective perturbation strength, whereas the thermalization time of the FPUT-$\beta$ chain remains essentially insensitive to the mass difference \cite{Feng_2022}.

\end{revision}

\begin{revision}These results show that inconsistent perturbation parametrizations can account for part of the diversity of reported exponents. They do not exclude genuinely different preasymptotic mechanisms associated with initial conditions, finite size, symmetry, hydrodynamic correlations, or the chosen equilibration observable.
\end{revision}

\begin{revision}
An inverse-square scaling shared by several models does not imply that their thermalization pathways are identical. In the cases compared here, the symmetry of the interaction potential can substantially modify the intermediate dynamics even when the same scaling of the final thermalization time is observed.
For instance, asymmetric interaction potentials enhance the contribution of higher-order processes \cite{e27070741} and induce asymmetric broadening of phonon spectral peaks \cite{Fu:2019R}. Studies of diatomic chains further show that the order in which equipartition is established between the acoustic and optical branches may even be reversed depending on the underlying mechanism, although the final thermalization time $T_{\mathrm{eq}}$ still follows the same scaling law \cite{PhysRevE.100.052102}. These results indicate that while the scaling of the thermalization time can be common across the models and regimes studied, the microscopic pathways toward equilibrium can vary substantially depending on the specific interaction mechanisms.
Moreover, recent studies of one-dimensional diatomic gases indicate that the inverse-square scaling of the thermalization time can persist even when the Lyapunov exponent vanishes, provided the system remains close to the integrable regime. When the system moves far away from this regime, long-time correlations associated with hydrodynamic effects may dominate the dynamics and modify the relaxation behavior. This crossover may also provide a possible perspective on the pronounced finite-size effects observed in lattice systems with symmetric interaction potentials \cite{fu2026DHPG}.
\end{revision}

\begin{figure}
    \centering
    \includegraphics[width=.7\columnwidth]{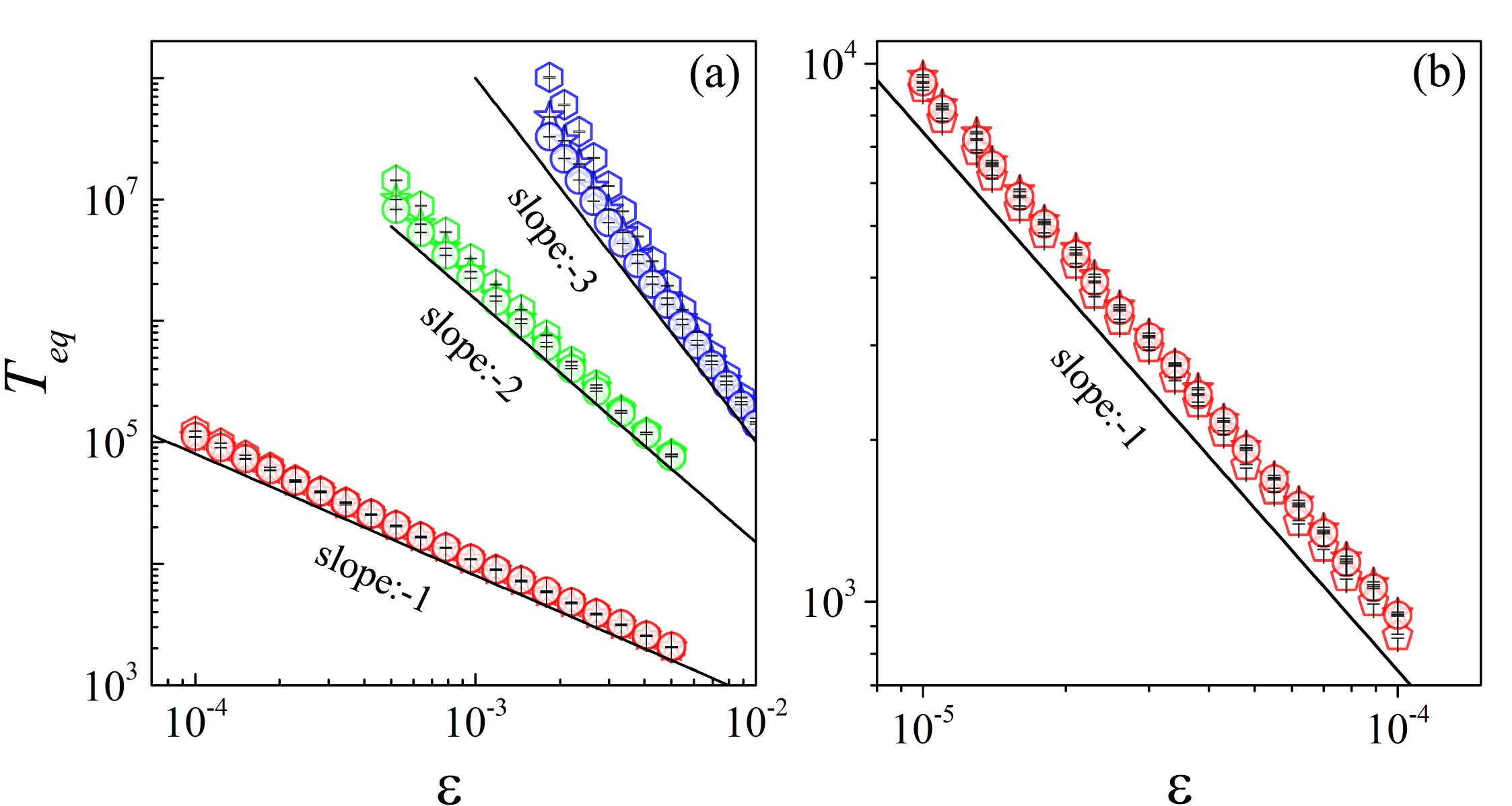}
    \caption{\rev{Thermalization scaling in one-dimensional mass-disordered lattices with fixed boundary conditions. The masses are chosen independently and identically from a uniform distribution over $[1-\delta m,1+\delta m]$, where the disorder strength is $\delta m=0.2$. (a) Polynomial interactions of orders $n=3,4,5$ (bottom to top) and (b) Lennard--Jones interactions, for $N=511$ (hexagons), $1023$ (stars), and $2047$ (circles). The lowest $10\%$ of the modes are initially excited, $T_{\mathrm{eq}}$ is defined by $\xi(T_{\mathrm{eq}})=0.5$, and 120 random initial states are averaged for each fixed disorder realization. The dashed lines show the expected large-$N$ energy-density slopes. The source data do not report a separate disorder-to-disorder variance for these plotted points. The error bars are much smaller than the symbols. Reprinted from Ref.~\cite{Wang:2020}.}}\label{fig:scaling_disorder1d_prl}
\end{figure}

\begin{revision}
Within the same theoretical framework, the stability of Anderson-localized phonon modes can also be analyzed from the perspective of wave resonances. For the one-dimensional models and regimes examined, nonlinear interactions can destabilize localized modes, and weak mass disorder can accelerate thermalization by relaxing wave-vector selection constraints and opening additional resonant or quasi-resonant channels \cite{Wang:2020,Sun:2020}. These conclusions are model- and regime-dependent and do not imply that disorder generically accelerates thermalization.
\end{revision}

\begin{revision}
Figure~\ref{fig:scaling_disorder1d_prl} shows the thermalization time of one-dimensional disordered lattices as a function of the energy density. For polynomial interactions of orders $n=3$, $4$, and $5$ [see again Eq.~(\ref{eq:Hnprime})], the numerical results progressively approach the kinetic prediction
$T_{\mathrm{eq}}\sim\varepsilon^{-(n-2)}$
as the system size is increased. At a fixed size, deviations emerge at sufficiently low energy densities, indicating that increasingly large systems are required to reach the kinetic regime as the effective nonlinearity decreases. For the disordered Lennard--Jones (LJ) lattice with the form of potential $V(x)=[1/(1+x)^6-1]^2/72$, $T_{\mathrm{eq}}\sim\varepsilon^{-1}$ because the cubic term is the leading nonlinearity in the low-energy expansion. This differs from the homogeneous counterpart, where three-wave resonances are forbidden and $T_{\mathrm{eq}}\sim\varepsilon^{-2}$, demonstrating that disorder can activate lower-order resonant channels and thereby accelerate thermalization \cite{Wang:2020}.

From the perspective of the nearest integrable reference Hamiltonian, the homogeneous LJ chain is closer to the Toda lattice than to the harmonic chain \cite{Benettin2023}. After choosing a Toda potential whose quadratic and cubic coefficients match those of the LJ potential, the residual nonintegrable perturbation begins at quartic order, accounting for the scaling $T_{\mathrm{eq}}\sim\varepsilon^{-2}$. Mass disorder, however, destroys the Toda integrability. For the mass-disordered LJ chain, the appropriate integrable reference system is therefore the harmonic Hamiltonian with the same mass configuration. Relative to this reference, the leading perturbation in the Taylor expansion of the LJ potential is cubic, yielding $T_{\mathrm{eq}}\sim\varepsilon^{-1}$. Thus, disorder changes not only the resonance structure but also the appropriate integrable reference point and the leading order of the effective nonintegrable perturbation.
\end{revision}

From an application perspective, real condensed-matter systems are primarily two- and three-dimensional. Therefore, extending the analysis from one dimension to higher dimensions is not merely a formal generalization but a necessary step toward understanding realistic material systems. However, the theoretical and numerical complexity increases significantly in higher-dimensional systems, and systematic results have long been lacking. Early pioneering works in the 1970s and 1980s already identified chaos thresholds in two-dimensional FPUT models \cite{Ooyama1969,Bocchieri1974,Benettin1980,Benettin1983}. Due to the limited computing power available at that time, important questions such as the scaling of the threshold with system size, the persistence of weak-chaos regimes, and the ultimate fate of FPUT recurrences remained unresolved.
With the advent of modern high-performance computing, Benettin and collaborators studied a two-dimensional triangular lattice and reported a power-law dependence of the thermalization time on the energy density. Interestingly, the exponent was found to depend on boundary conditions: $\gamma=1$ for fixed or open boundaries and $\gamma=5/4$ for periodic boundaries \cite{benettin_time_2005,Benettin:2008}. Meanwhile, studies motivated by realistic materials such as graphene have shown that out-of-plane flexural modes play a crucial role in energy redistribution and thermal transport \cite{Midtvedt2014,Wang2018,Wang2018a,Deng_2026}.

\begin{figure}[t]
    \centering
    \includegraphics[width=.7\columnwidth]{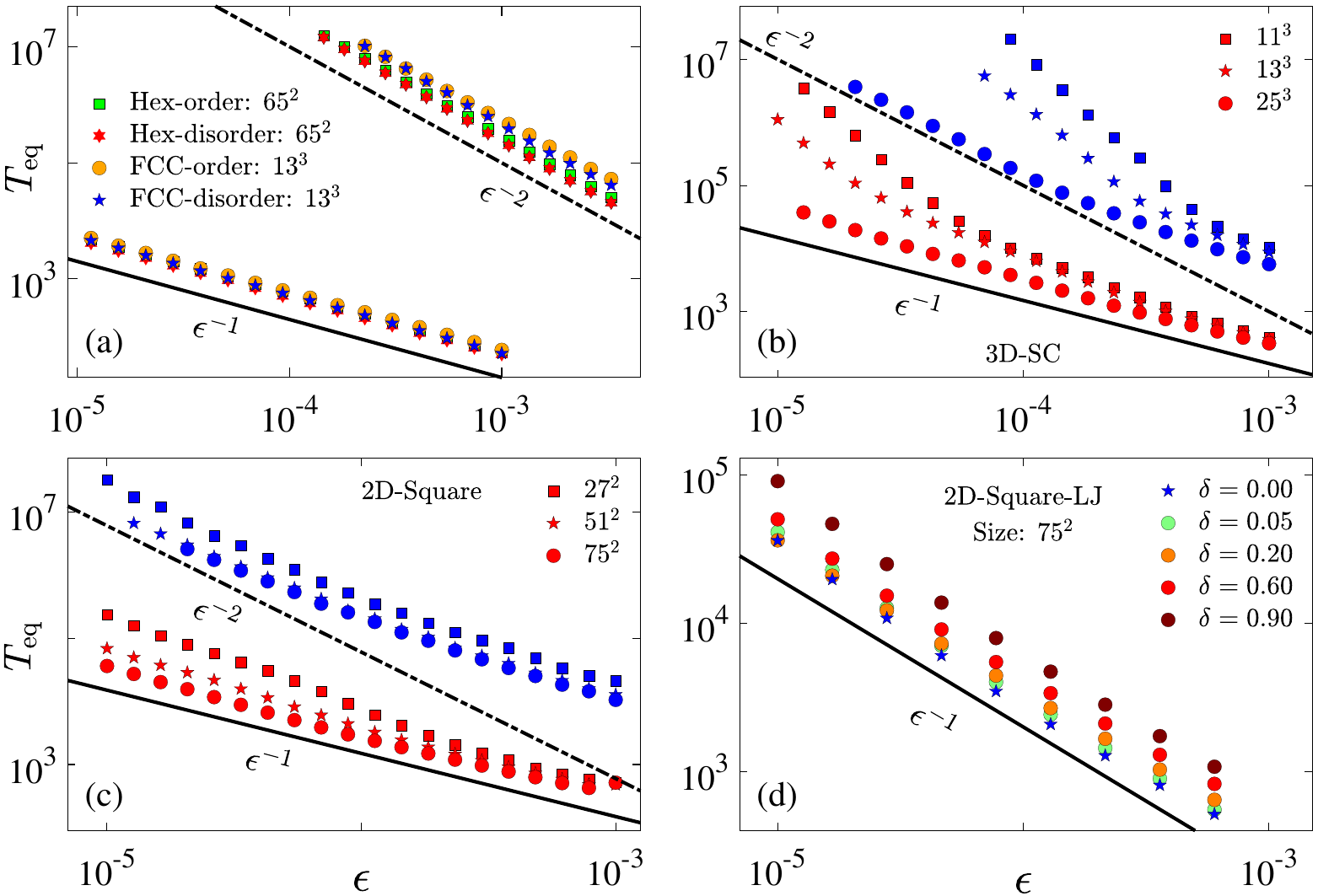}
    \caption{\rev{Thermalization time versus energy density in higher-dimensional lattices, all simulated with fixed boundary conditions and the lowest $10\%$ of modes initially excited. The results were averaged over 20 random-phase realizations. (a) Ordered and disordered hexagonal and fcc lattices in sizes beyond the strongest finite-size regime, and $65^2$ denotes the system size being $65\times65$. (b) Simple-cubic lattices: blue squares, black stars, and red circles denote system size $11\times11\times11$, $13\times13\times13$, and $25\times25\times25$, respectively. (c) Square lattices: blue squares, black stars, and red circles denote $27^2$, $51^2$, and $75^2$, respectively; hence the blue symbols questioned in the report are the smallest systems. (d) Square-lattice results at different disorder strengths for size $75^2$. Dashed lines are theoretical reference slopes. Fits reported in Ref.~\cite{Wang:2024} give $T_{\mathrm{eq}}\sim\varepsilon^{-1\pm0.04}$ for the Lennard--Jones potential and $T_{\mathrm{eq}}\sim\varepsilon^{-2\pm0.05}$ for the quartic potential in panel (a); at the largest square size the corresponding low-energy LJ slope is $-1\pm0.10$, whereas the quartic data have only reached an effective slope near $-1.7$ at the computational limit. The error bars (not shown) are much smaller than the symbols. Reprinted from Ref.~\cite{Wang:2024}.}}\label{fig:Teq_highD}
\end{figure}

\begin{revision}
Despite these pioneering efforts, our current understanding of thermalization in higher-dimensional lattices remains relatively limited. Existing studies are largely restricted to specific lattice geometries (such as triangular lattices) and particular interaction potentials (e.g., Lennard--Jones potentials). Consequently, the influence of spatial dimensionality on the qualitative and quantitative features of thermalization, as well as the role of disorder in higher-dimensional lattices, remains insufficiently understood.
\end{revision}
\begin{revision}
In recent years, the Xiamen University group has investigated two- and three-dimensional lattice systems. For the geometries, interaction potentials, system sizes, and energy ranges examined, the results are consistent with the inverse-square thermalization scaling (see Fig.~\ref{fig:Teq_highD}) and also show that spatial dimensionality affects the relaxation process. Higher-dimensional lattices have larger coordination numbers and can exhibit symmetry-induced mode degeneracies, both of which modify the available channels for resonant energy transfer. The numerical results indicate that these structural features can alter the density and organization of resonance couplings; their precise effect remains dependent on lattice geometry, boundary conditions, and disorder.
\end{revision}

\begin{revision}

Figure~\ref{fig:Teq_highD} summarizes the thermalization behavior of representative two- and three-dimensional lattices under fixed boundary conditions, with the lowest-frequency $10\%$ of the normal modes initially excited. Figure~\ref{fig:Teq_highD}(a) shows that both homogeneous and mass-disordered hexagonal and fcc lattices closely follow the kinetic predictions
$T_{\mathrm{eq}}\sim\varepsilon^{-1}$ for the Lennard--Jones potential and
$T_{\mathrm{eq}}\sim\varepsilon^{-2}$ for the quartic potential, with fitted exponents
$-1\pm0.04$ and $-2\pm0.05$, respectively. In these relatively low-degeneracy lattices, weak mass disorder slightly accelerates thermalization by relaxing the resonance constraints.
Figures~\ref{fig:Teq_highD}(b) and \ref{fig:Teq_highD}(c) present the corresponding results for homogeneous simple-cubic and square lattices. The predicted scalings are gradually recovered as the system size is increased and the energy density is reduced, but the convergence is substantially slower than in the fcc and hexagonal lattices. This strong finite-size dependence originates from the highly degenerate spectra of the simple-cubic and square lattices, for which many modes share the same frequency and therefore constitute only a smaller number of distinct spectral levels. At the largest accessible sizes, the Lennard--Jones models approach the predicted $\varepsilon^{-1}$ scaling, whereas the quartic square lattice still exhibits an effective exponent of approximately $-1.7$, indicating that substantially larger systems are required to reach the asymptotic $\varepsilon^{-2}$ regime. Figure~\ref{fig:Teq_highD}(d) further illustrates the effect of mass disorder in the square lattice. In contrast to the weak acceleration observed in the low-degeneracy hexagonal and fcc lattices, increasing the disorder strength systematically prolongs the thermalization time while leaving its approximate energy-density scaling unchanged. This opposite response arises because disorder lifts the extensive frequency degeneracy of the square lattice and thereby suppresses the rapid resonant energy exchange among initially degenerate modes. Thus, disorder may accelerate thermalization in low-degeneracy lattices by enlarging the available quasi-resonant phase space, but retard it in highly degenerate lattices by removing an efficient intra-degeneracy energy-transfer channel.

Taken together, these results highlight the combined roles of dimensionality, spectral degeneracy, and resonance-network connectivity in controlling the efficiency of energy redistribution and the finite-size approach to kinetic thermalization in nonlinear lattice systems.
\end{revision}

\section{Thermalization classes}\label{sec4}

The results presented in the previous sections might suggest that sufficiently large nonlinear lattices always thermalize according to the inverse-square scaling law. However, this inference would extend beyond the models and regimes examined so far.
In this section, following Ref.~\cite{Lin:2025}, we discuss a proposed organization of the models studied into two thermalization classes according to the eigenmodes of the reference integrable Hamiltonian. When the reference spectrum contains extended modes, the numerical results are consistent with an inverse-square thermalization law in sufficiently large systems; we refer to these cases as the first thermalization class. When all reference modes are localized, the models studied display a qualitatively different sequence of finite-range scaling regimes,
\begin{equation}
T_{\mathrm{eq}}\sim g^{-m}, \qquad m=2,4,6,\ldots ,
\end{equation}
where higher effective exponents appear as $g$ decreases over the accessible parameter ranges. The data indicate a strong suppression of thermalization and a much weaker size dependence than in systems with extended modes, but they do not establish a nonzero perturbation threshold or the absence of thermalization in the infinite-time limit. We refer to these cases as the second thermalization class. This classification is a framework motivated by resonance-network analysis and numerical evidence for the models considered, rather than an exhaustive theorem for generic classical lattices.


\rev{To illustrate the qualitative difference between the two thermalization classes defined above within the present resonance-network framework, we consider the $\phi^4$ lattice model \cite{Kevrekidis:2019}, whose Hamiltonian is}
\begin{equation}\label{eq:H}
H = \sum_j \left[
\frac{p_j^2}{2m_j}
+
\frac{1}{2}(x_{j+1}-x_j)^2
+
\frac{1}{2} b x_j^2
+
\frac{1}{4}\beta x_j^4
\right].
\end{equation}
Here, each mass $m_j$ is drawn independently from a uniform distribution over $[1-\delta m,\,1+\delta m]$, where $\delta m$ characterizes the disorder strength, while $b$ controls the amplitude of the on-site potential. Because of the presence of the on-site potential, this model possesses an important advantage: by tuning system parameters, the eigenmodes of the corresponding integrable Hamiltonian (obtained by removing the quartic term) can be made fully extended, partially extended with coexistence of localized modes, or completely localized. Therefore, this model provides a convenient setting in which to compare the two proposed thermalization classes.

\begin{revision}
Rescaling the coordinates and momenta using the energy density $\varepsilon$ (i.e., $x_j=\tilde x_j \varepsilon^{1/2}$ and $p_j=\tilde p_j \varepsilon^{1/2}$), the Hamiltonian can be written as
\begin{equation}
\tilde H = H/\varepsilon = H_0(\tilde x_j,\tilde p_j) + \sum_j \frac{1}{4} g\, \tilde x_j^4 ,
\end{equation}
where $g = \beta \varepsilon$ characterizes the effective nonlinear (i.e., nonintegrability) strength of the system.
\end{revision}

\begin{figure*}[htb]
\centering
\includegraphics[width=.9\columnwidth]{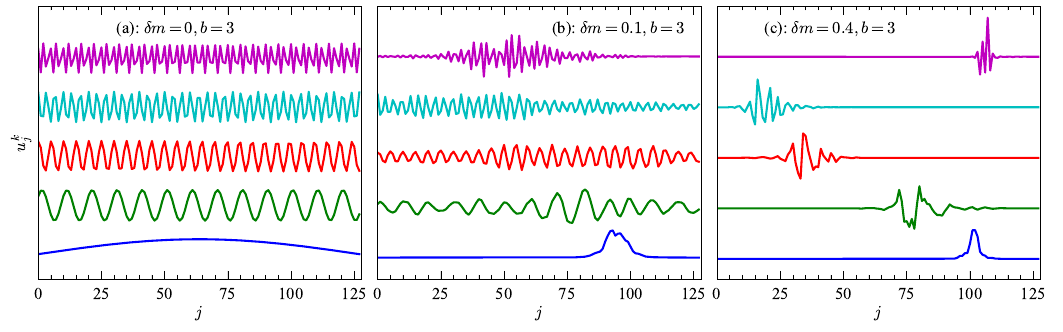}
\caption{Eigenmodes $u_j^k$ of the lattice for $N=128$ and $b=3$ under different disorder strengths:
(a) ordered lattice ($\delta m=0$), (b) weak disorder, and (c) strong disorder.
As the disorder increases, the eigenmodes evolve from extended states to localized states. From bottom to top, the eigenfrequencies $\omega_k$ increase.}\label{fig:ujk}
\end{figure*}

As illustrated in Fig.~\ref{fig:ujk}, depending on the system parameters the eigenmodes of the lattice may appear as fully extended states, Fig.~\ref{fig:ujk}(a); coexistence of extended and localized modes, Fig.~\ref{fig:ujk}(b); or completely localized modes, Fig.~\ref{fig:ujk}(c). For the one-dimensional $\phi^4$ model with independent mass disorder considered here, the ordered lattice has extended reference modes, whereas at fixed nonzero disorder the modes become localized in the thermodynamic limit. The apparent coexistence of extended and localized modes is a finite-size regime whose extent depends on the disorder strength \cite{Lin:2025}.

Based on perturbation theory, the kinetic equation for the eigenmodes of the $\phi^4$ lattice can be written as
\begin{equation}\label{eq:4keq}
\partial_t \langle I_{k_1} \rangle = -2 g^2 \pi \sum_{k_2,k_3,k_4} A^2_{k_1k_2k_3k_4}
\Big(
C_{k_1k_2k_3k_4}\Delta_{k_1k_2k_3k_4}
+ 9C_{k_1k_2k_3}^{k_4}\Delta_{k_1k_2k_3}^{k_4}
+ 9C^{k_3k_4}_{k_1k_2}\Delta_{k_1k_2}^{k_3k_4}
 + C^{k_2k_3k_4}_{k_1}\Delta_{k_1}^{k_2k_3k_4}
\Big),
\end{equation}
where
\begin{equation}
A_{k_1k_2k_3k_4}
=
\frac{1}{4\sqrt{\omega_{k_1}\omega_{k_2}\omega_{k_3}\omega_{k_4}}}
\sum_j
\frac{u^{k_1}_j u^{k_2}_j u^{k_3}_j u^{k_4}_j}{m_j^2}
\end{equation}
denotes the overlap integral of four eigenmodes. The coefficients inside the parentheses have the form
\begin{equation}
C^{k_3k_4}_{k_1k_2}
=
\langle I_{k_1}  \rangle \langle I_{k_2}\rangle  \langle I_{k_3} \rangle \langle I_{k_4} \rangle
\left(
\frac{1}{\langle I_{k_3} \rangle }+\frac{1}{\langle I_{k_4} \rangle }
-\frac{1}{\langle I_{k_1}  \rangle }-\frac{1}{\langle I_{k_2}\rangle  }
\right),
\end{equation}
and
\begin{equation}
\Delta_{k_1k_2}^{k_3k_4}
=
\delta(\omega_{k_3}+\omega_{k_4}-\omega_{k_1}-\omega_{k_2}),
\end{equation}
while the other terms are defined analogously.
\begin{revision}
Under the kinetic closure, Eq.~\eqref{eq:4keq} can be reduced to the form of Eq.~\eqref{eq:kenetic_equation}. As long as $\gamma_{k_1}$ remains nonzero, the relaxation time of mode $k_1$ follows $\gamma_{k_1}^{-1}\propto g^{-2}$. Relating this modal relaxation rate to the global thermalization time additionally requires the kinetic closure to remain valid and the relevant resonance network to connect the modes needed for global energy redistribution. For a given disorder strength, these conditions may fail as the nonlinear parameter decreases; in the fully localized cases studied, the observed crossover has only a weak dependence on system size.
\end{revision}

To understand this phenomenon, a quantitative quasi-resonance criterion is useful. Quasi-resonance analysis has been widely used and its importance in energy diffusion and transport processes has been well recognized \cite{Pushkarev:1999,Connaughton:2001,Kartashova:2007,Lvov:2010,Pan:2017,Wang:2024}. Nevertheless, conditions similar to Eq.~\eqref{eq:quasi-MWRC_w} have previously been applied mainly in a qualitative manner.
Here we supplement the exact resonance condition with the quasi-resonant criterion
\begin{equation}\label{eq:zhun}
\left| \omega_{k_1} \pm \omega_{k_2} \pm \omega_{k_3} \pm \omega_{k_4} \right| < \Omega
\quad \& \quad
A_{k_1k_2k_3k_4} \ne 0 ,
\end{equation}
which provides the basis for the quantitative quasi-resonance analysis used here. The parameter $\Omega$ characterizes the nonlinear frequency broadening; larger nonlinearities correspond to larger $\Omega$, and vice versa.

To characterize the local connectivity of the resonance network, we further introduce the connectivity proxy
\begin{equation}\label{eq:p4}
p_4(k_1)=
\sum_{k_2,k_3,k_4}
\left|A_{k_1k_2k_3k_4}\right|,
\end{equation}
which measures the total overlap weight of the quasi-resonant combinations involving mode $k_1$. Here $k_2,k_3,k_4$ run over all mode combinations satisfying the quasi-resonance condition \eqref{eq:zhun}. If $p_4(k_1)=0$, the corresponding mode has no link within this four-wave criterion. Larger values of $p_4(k_1)$ indicate stronger local coupling to other modes, but neither $p_4(k_1)>0$ for every mode nor a large average $\langle p_4\rangle$ alone proves that the full graph is globally connected.
When the quasi-resonant graph is globally connected and the other kinetic assumptions hold, energy can spread across mode space and the relaxation dynamics are expected to be consistent with $T_{\mathrm{eq}}\propto g^{-2}$. The quantities $p_4(k_1)$ and $\langle p_4(k_1)\rangle$ are therefore used below as indicators of how the resonance network changes as the perturbation is reduced. Within the models studied, this analysis motivates the distinction between the two proposed thermalization classes.

\begin{figure}[tb]
\centering
\includegraphics{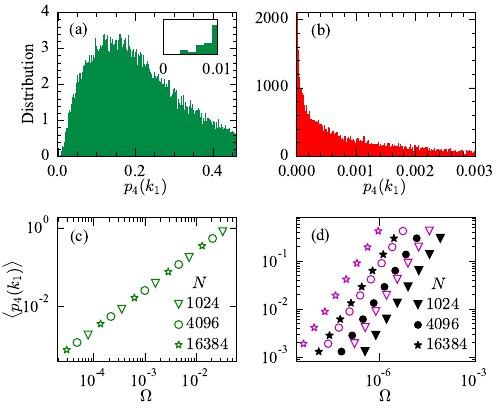}
\caption{\rev{
Local-connectivity proxy for the resonance network characterized by $p_4(k_1)$.
(a) and (b) Distribution of $p_4(k_1)$ for parameters $b=3$, $\delta m=0.5$, and $N=16384$ at two values of the quasi-resonance width: (a) $\Omega=0.01$ and (b) $\Omega=0.0001$. The inset in (a) shows a magnified view near the origin.
(c) Dependence of the average connectivity $\langle p_4(k_1)\rangle$ on $\Omega$ for a lattice with completely localized modes ($b=3$, $\delta m=0.5$).
(d) $\langle p_4(k_1)\rangle$ as a function of $\Omega$ for lattices with fully extended modes ($b=3$, $\delta m=0$, solid symbols) and for lattices with coexistence of extended and localized modes ($b=0.1$, $\delta m=0.05$, open symbols). Reprinted from Ref. \cite{Lin:2025}.}
}
\label{fig:zhun}
\end{figure}

The distribution of $p_4(k_1)$ provides important information about the internal connectivity of the resonance network. Figures~\ref{fig:zhun}(a) and \ref{fig:zhun}(b) display the distributions of $p_4(k_1)$ for a lattice with completely localized modes under two values of $\Omega$, corresponding respectively to relatively strong and weak nonlinear perturbations.
When $\Omega=0.01$, the distribution of $p_4(k_1)$ exhibits a clear gap near the origin, indicating that every mode has at least one link within the four-wave criterion and that most modes have relatively large local-connectivity weights. In contrast, when $\Omega=10^{-4}$ the distribution develops a power-law-like form and a sharp peak appears near the origin. This indicates that many modes have very weak or vanishing four-wave links and is consistent with increasing fragmentation of the resonance network.
These results suggest that, in the fully localized cases studied, the four-wave resonance network becomes progressively less connected as $\Omega$ decreases. The four-wave kinetic description and its $g^{-2}$ scaling may then cease to control the observed relaxation. Higher-order processes, such as six-wave and eight-wave interactions, provide candidate mechanisms for the subsequent regimes. Extending the connectivity analysis to these processes suggests successive finite-range scalings of the form $g^{-4}$, $g^{-6}$, and so on; the persistence of this hierarchy in the asymptotic weak-coupling limit remains an open question.

To further quantify the connectivity of the resonance network, we compute the average connectivity strength
\begin{equation}
\langle p_4(k_1)\rangle=\frac{1}{N}\sum_{k_1} p_4(k_1).
\end{equation}
Figure~\ref{fig:zhun}(c) shows $\langle p_4(k_1)\rangle$ as a function of $\Omega$ for three different system sizes in a lattice with completely localized modes. The data for different system sizes collapse onto a single curve, indicating that the connectivity strength in fully localized lattices is essentially independent of system size. The observed power-law behavior implies that $\langle p_4(k_1)\rangle$ becomes progressively smaller as the nonlinear strength decreases, eventually leading to a loss of connectivity of the four-wave quasi-resonant network. The fragmentation of the network is directly illustrated by the distributions shown in Figs.~\ref{fig:zhun}(a) and \ref{fig:zhun}(b).

For comparison, Fig.~\ref{fig:zhun}(d) shows the corresponding results for lattices containing extended modes. Each data set again includes three system sizes. For a fixed system size, $\langle p_4(k_1)\rangle$ decreases monotonically as $\Omega$ decreases, indicating that the four-wave quasi-resonant network would eventually break down at sufficiently weak perturbations. However, in contrast to the fully localized case, we observe that for fixed $\Omega$ the average connectivity $\langle p_4(k_1)\rangle$ increases monotonically with system size. This implies that in sufficiently large lattices the connectivity can become arbitrarily large, ensuring that the four-wave quasi-resonant network remains globally connected. As a result, the universal scaling $T_{\mathrm{eq}}\propto g^{-2}$ continues to govern the thermalization dynamics in the large-system limit.

\begin{figure}[tp]
\centering
\includegraphics{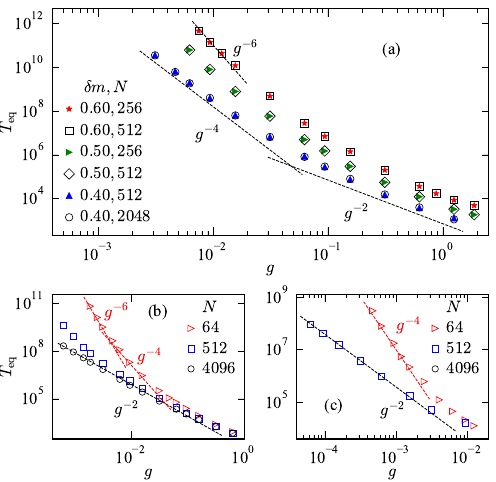}
\caption{
\rev{Thermalization time versus nonlinear strength $g$ for the $\phi^4$ lattice. (a) Fully localized cases, with each symbol in the legend specifying the pair $(\delta m,N)$. (b) Fully extended modes ($b=3$, $\delta m=0$) and (c) mixed modes ($b=0.1$, $\delta m=0.05$), each shown for $N=64,512,4096$. The dashed lines are reference slopes $g^{-2}$, $g^{-4}$, and $g^{-6}$, not global regression fits. The error bars (not shown) are much smaller than the symbols. Reprinted from Ref.~\cite{Lin:2025}.}
}
\label{fig:phi4}
\end{figure}

Direct numerical simulations confirm these theoretical predictions. The dependence of the thermalization time on the nonlinear strength in fully localized lattices is shown in Fig.~\ref{fig:phi4}(a). For fixed values of $\delta m$, $b$, and $g$, the thermalization times obtained for different system sizes are nearly identical, indicating that the relaxation time is essentially independent of system size. As $g$ decreases, the scaling of the thermalization time gradually changes from the four-wave resonance regime to those associated with six-wave and eight-wave resonances.
It is worth noting that for disorder strength $\delta m=0.4$, the lattice with $N=256$ still contains extended modes. Therefore, a larger system size is used in the analysis, although the resulting thermalization times remain nearly identical for different sizes. Combining the trends shown in Fig.~\ref{fig:zhun}(c) with the disorder dependence observed in Fig.~\ref{fig:phi4}(a), we infer that as $g$ decreases further, the thermalization dynamics of fully localized systems generally undergo successive transitions from lower-order to higher-order resonance scaling regimes. In other words, fully localized lattices possess an effective nonlinear threshold for thermalization, which decreases as the nonlinear strength becomes weaker.
This mechanism ultimately distinguishes the second universality class of thermalization, in which the resonance network fragments as the perturbation strength decreases.

Figures~\ref{fig:phi4}(b) and \ref{fig:phi4}(c) show the corresponding results for lattices with fully extended modes and mixed-mode lattices (containing both extended and localized modes), respectively. In both cases, small systems exhibit a crossover from the thermalization scaling governed by four-wave resonances to those dominated by higher-order resonances as $g$ decreases.
In contrast, for sufficiently large systems the numerical results remain consistent with the prediction of four-wave resonance theory over the entire range of parameters considered.
This size dependence reflects the persistence of a globally connected four-wave resonance network in lattices containing extended modes.

\section{Conclusion and Discussion}\label{sec5}

In summary, on the occasion of the 70th anniversary of the FPUT problem, a coherent picture of the thermalization laws of weakly nonlinear lattices in the thermodynamic limit has begun to emerge. In the regime of weak effective perturbations, perturbative analysis can lead to a kinetic description of the evolution of normal-mode energies and predict an inverse-square dependence of the thermalization time on the effective nonlinear strength, provided that the leading relaxation process is nonvanishing and the assumptions underlying the kinetic description are satisfied. Specifically, if the Hamiltonian is written as $H=H_0+gV$, where $H_0$ represents the integrable reference system and $g$ characterizes the effective perturbation strength, the resulting scaling is
\begin{equation*}
T_{\mathrm{eq}}\propto g^{-2}.
\end{equation*}
Further examination of the validity conditions of the kinetic equation deepens our understanding of lattice thermalization.

\begin{revision}
Within this framework, the modes relevant to global energy redistribution need to be sufficiently connected through resonant or quasi-resonant $n$-wave processes, so that energy can propagate across the mode space. The investigation of this connectivity suggests different thermalization classes.
For the first class, in which the eigenmodes of the reference system are extended, the lowest effective nonlinear interaction in the Taylor expansion of the interaction potential can generate a connected $n$-wave resonance network in the thermodynamic limit. Consequently, the inverse-square scaling $T_{\mathrm{eq}}\propto g^{-2}$ has been observed across a broad range of the lattice models studied so far. Numerical simulations indicate that, in the cases considered, this scaling can already appear clearly in systems containing only several thousand particles. An important point is that its observation requires an appropriate decomposition of the Hamiltonian: the system should be separated into a suitable integrable part and a perturbation around that integrable point, so that the effective nonlinear parameter correctly measures the deviation from integrability. For perturbed-Toda systems, resonant-interaction analyses provide a wave-turbulence rationale for relaxation when the on-shell scattering cancellations of the integrable Toda lattice are broken \cite{Onorato:2015,Onorato:2023}; however, an explicit kinetic equation formulated in terms of Toda actions or quasiparticle distributions, and capable of relating the resulting collision rate to the global thermalization time, remains to be established. Some quantitative scalings reported for systems within this broad class have also not yet been fully reconciled.

It is worth noting that, for one-dimensional lattices, the scaling $T_{\mathrm{eq}}\propto \varepsilon^{-9/4}$ reported for the FPUT-$\alpha+\beta$ chain in Ref.~\cite{Benettin:2011} differs from the inverse-square behavior reviewed here for perturbed Toda systems. Besides the finite-size effect and higher-order resonance interactions causing such deviations, we speculate that this discrepancy arises because the small deviation of the system from the nearest exactly integrable point was not fully incorporated into the effective measure of nonintegrability. When the energy density is used directly to characterize the perturbation, such a deviation may produce a noninteger effective exponent that depends on the finite parameter range examined, rather than indicating a distinct asymptotic universality class. The two-dimensional case requires separate consideration. The scaling obtained here under fixed boundary conditions differs from the relation $T_{\mathrm{eq}}\propto \varepsilon^{-5/4}$
reported under periodic boundary conditions in Refs.~\cite{benettin_time_2005,Benettin:2008}. Because boundary conditions modify the spectral sampling, mode degeneracies, and available resonance channels, a more systematic finite-size analysis is required to determine whether this discrepancy is primarily a finite-size effect or reflects genuinely different asymptotic behavior under different boundary conditions.

While these comparisons concern the precise scaling within the first class, a qualitatively different behavior arises in the second class, in which the eigenmodes of the reference system are completely localized. In the models studied to date, the connectivity of the resonance network decreases as the effective nonlinear strength is reduced and can eventually lead to fragmentation of the low-order $n$-wave network. Over the accessible parameter and observation-time ranges, the thermalization time exhibits a hierarchy of scaling regimes,
\begin{equation*}
T_{\mathrm{eq}}\propto g^{-m}, \qquad m=2,4,6,\ldots,
\end{equation*}
with higher effective exponents appearing as the nonlinear strength becomes weaker. This behavior indicates a strong suppression of thermalization by weak perturbations and raises the possibility of effective thermal-insulating behavior. Another notable feature found in the models and parameter ranges studied is that the thermalization law has a much weaker dependence on system size than in systems with extended modes. Whether the exponent hierarchy persists asymptotically, whether arbitrarily small perturbations eventually lead to thermalization at sufficiently long times, and whether a genuine thermal-insulating state exists remain open questions.

The investigation of resonance-network connectivity also raises the long-standing question of whether finite lattices can always thermalize. Based on exact resonance conditions, it has been shown that six-wave resonances under periodic boundary conditions can form a connected network \cite{Onorato:2015}, suggesting that the finite systems considered there may ultimately thermalize with a power law scaling. However, our recent studies indicate that a kinetic description based only on exact resonances may become insufficient during the evolution because of intrinsic symmetry constraints. Instead, quasi-resonances appear to dominate energy diffusion in the regimes studied. Since the strength of quasi-resonances and the connectivity of the corresponding resonance network depend on the effective nonlinear parameter, the thermalization time can display a hierarchy $T_{\mathrm{eq}}\propto g^{-m}$, with the effective exponent $m$ increasing as the nonlinearity decreases. This behavior suggests the possibility of an effective nonlinear threshold over finite observation times, while the existence of a genuine nonzero threshold in the infinite-time limit remains to be established. Clarifying this issue is important not only for revisiting the original FPUT experiment but also for establishing the statistical-mechanical foundations of finite systems.
\end{revision}

Despite these important advances, the FPUT problem remains far from fully resolved. Seventy years after its discovery, it continues to open new directions of research. From a theoretical perspective, several important problems remain. First, while the weakly nonlinear regime is better characterized within the perturbative framework, thermalization in the strongly nonlinear or strongly nonintegrable regime remains poorly understood. Numerical simulations indicate systematic deviations from the inverse-square scaling in this regime, implying that perturbative kinetic theory is no longer sufficient and that new theoretical approaches are required. A recent study of a one-dimensional diatomic gas found a crossover from inverse-square kinetic relaxation near integrability to hydrodynamically dominated relaxation far from integrability \cite{fu2026DHPG}. Although this model is not a lattice, the crossover provides a candidate picture for investigating thermalization mechanisms in strongly nonintegrable lattices. Strong nonlinearity can also lead to rich phenomena, including many-body-localization-like behaviors in classical lattices \cite{campbell2004localizing,RevModPhys.91.021001}.
Second, thermalization and thermal transport are believed to be closely related at the microscopic level. Previous studies have revealed relations between the scaling exponents governing diffusion and anomalous heat conduction in particular models \cite{PhysRevLett.91.044301}. Recently, we have established such connections systematically in a one-dimensional diatomic gas model \cite{fu2026DHPG}. Further exploration of the generality of this relationship is important because heat conduction is one of the fundamental macroscopic properties of materials.
Third, the study of high-dimensional lattices has only just begun. The inverse-square thermalization scaling has been observed in the higher-dimensional lattices and parameter ranges studied so far, but the dynamical processes in higher dimensions are considerably richer. The spatial structure of higher-dimensional lattices introduces additional effects such as mode degeneracy, interactions between acoustic and optical phonon branches, and disorder-induced localization, all of which can significantly influence energy-transfer processes in materials. Understanding the roles of different multiphonon processes, such as three-phonon and four-phonon interactions, therefore provides promising directions for future research.

More broadly, the thermalization of open systems also deserves extensive investigation. Within such a framework, the role of integrability in determining the relaxation of a system toward equilibrium should be reconsidered from new perspectives \cite{e26020156}. Finally, exploring the similarities and differences between thermalization in classical and quantum systems represents another important direction for future studies.

\bmhead{Conflicts of Interest} The authors declare no conflicts of interest regarding this manuscript.

\bmhead{Funding} This work was supported by the National Natural Science Foundation of China (Grants Nos. 12247106, 12465010, 12505052, 12575042). W. Fu also acknowledges support from the Long-yuan Youth Talents Project of Gansu Province, the Fei-tian Scholars Project of Gansu Province, the Leading Talent Project of Tianshui City, and the Innovation Fund from the Department of Education of Gansu Province (Grant No.~2023A-106).

\bmhead{Data availability}  Data are contained within the article.


\begin{thebibliography}{75}
\ifx \bisbn   \undefined \def \bisbn  #1{ISBN #1}\fi
\ifx \binits  \undefined \def \binits#1{#1}\fi
\ifx \bauthor  \undefined \def \bauthor#1{#1}\fi
\ifx \batitle  \undefined \def \batitle#1{#1}\fi
\ifx \bjtitle  \undefined \def \bjtitle#1{#1}\fi
\ifx \bvolume  \undefined \def \bvolume#1{\textbf{#1}}\fi
\ifx \byear  \undefined \def \byear#1{#1}\fi
\ifx \bissue  \undefined \def \bissue#1{#1}\fi
\ifx \bfpage  \undefined \def \bfpage#1{#1}\fi
\ifx \blpage  \undefined \def \blpage #1{#1}\fi
\ifx \burl  \undefined \def \burl#1{\textsf{#1}}\fi
\ifx \doiurl  \undefined \def \doiurl#1{\url{https://doi.org/#1}}\fi
\ifx \betal  \undefined \def \betal{\textit{et al.}}\fi
\ifx \binstitute  \undefined \def \binstitute#1{#1}\fi
\ifx \binstitutionaled  \undefined \def \binstitutionaled#1{#1}\fi
\ifx \bctitle  \undefined \def \bctitle#1{#1}\fi
\ifx \beditor  \undefined \def \beditor#1{#1}\fi
\ifx \bpublisher  \undefined \def \bpublisher#1{#1}\fi
\ifx \bbtitle  \undefined \def \bbtitle#1{#1}\fi
\ifx \bedition  \undefined \def \bedition#1{#1}\fi
\ifx \bseriesno  \undefined \def \bseriesno#1{#1}\fi
\ifx \blocation  \undefined \def \blocation#1{#1}\fi
\ifx \bsertitle  \undefined \def \bsertitle#1{#1}\fi
\ifx \bsnm \undefined \def \bsnm#1{#1}\fi
\ifx \bsuffix \undefined \def \bsuffix#1{#1}\fi
\ifx \bparticle \undefined \def \bparticle#1{#1}\fi
\ifx \barticle \undefined \def \barticle#1{#1}\fi
\bibcommenthead
\ifx \bconfdate \undefined \def \bconfdate #1{#1}\fi
\ifx \botherref \undefined \def \botherref #1{#1}\fi
\ifx \url \undefined \def \url#1{\textsf{#1}}\fi
\ifx \bchapter \undefined \def \bchapter#1{#1}\fi
\ifx \bbook \undefined \def \bbook#1{#1}\fi
\ifx \bcomment \undefined \def \bcomment#1{#1}\fi
\ifx \oauthor \undefined \def \oauthor#1{#1}\fi
\ifx \citeauthoryear \undefined \def \citeauthoryear#1{#1}\fi
\ifx \endbibitem  \undefined \def \endbibitem {}\fi
\ifx \bconflocation  \undefined \def \bconflocation#1{#1}\fi
\ifx \arxivurl  \undefined \def \arxivurl#1{\textsf{#1}}\fi
\csname PreBibitemsHook\endcsname

\bibitem[\protect\citeauthoryear{Fermi et~al.}{1955}]{Fermi:1955}
\begin{barticle}
\bauthor{\bsnm{Fermi}, \binits{E.}},
\bauthor{\bsnm{Pasta}, \binits{J.}},
\bauthor{\bsnm{Ulam}, \binits{S.}}:
\batitle{{Studies of the nonlinear problems}}.
\bjtitle{Los Alamos Scientific Laboratory, Report No. LA-1940}
(\byear{1955})
\doiurl{10.2172/4376203}
\end{barticle}
\endbibitem

\bibitem[\protect\citeauthoryear{Ford}{1992}]{Ford:1992}
\begin{barticle}
\bauthor{\bsnm{Ford}, \binits{J.}}:
\batitle{The fermi-pasta-ulam problem: Paradox turns discovery}.
\bjtitle{Phys. Rep.}
\bvolume{213}(\bissue{5}),
\bfpage{271}--\blpage{310}
(\byear{1992})
\doiurl{10.1016/0370-1573(92)90116-H}
\end{barticle}
\endbibitem

\bibitem[\protect\citeauthoryear{Campbell et~al.}{2005}]{Campbell:2005}
\begin{barticle}
\bauthor{\bsnm{Campbell}, \binits{D.K.}},
\bauthor{\bsnm{Rosenau}, \binits{P.}},
\bauthor{\bsnm{Zaslavsky}, \binits{G.M.}}:
\batitle{{Introduction: The Fermi--Pasta--Ulam problem---The first fifty
  years}}.
\bjtitle{Chaos}
\bvolume{15}(\bissue{1}),
\bfpage{015101}
(\byear{2005})
\doiurl{10.1063/1.1889345}
\end{barticle}
\endbibitem

\bibitem[\protect\citeauthoryear{Gallavotti}{2008}]{gallavotti_fermi-pasta-ulam_2008}
\begin{bbook}
\beditor{\bsnm{Gallavotti}, \binits{G.}} (ed.):
\bbtitle{The Fermi-Pasta-Ulam Problem: A Status Report}.
\bsertitle{Lecture Notes in Physics},
vol. \bseriesno{728}.
\bpublisher{Springer},
\blocation{Berlin, Heidelberg}
(\byear{2008}).
\doiurl{10.1007/978-3-540-72995-2}
\end{bbook}
\endbibitem

\bibitem[\protect\citeauthoryear{Kolmogorov}{1954}]{Kolmogorov1954}
\begin{barticle}
\bauthor{\bsnm{Kolmogorov}, \binits{A.N.}}:
\batitle{On conservation of conditionally periodic motions for a small change
  in {Hamilton}'s function}.
\bjtitle{Doklady Akademii Nauk SSSR}
\bvolume{98},
\bfpage{527}--\blpage{530}
(\byear{1954})
\end{barticle}
\endbibitem

\bibitem[\protect\citeauthoryear{Arnol'd}{1963}]{arnol1963small}
\begin{barticle}
\bauthor{\bsnm{Arnol'd}, \binits{V.I.}}:
\batitle{Small denominators and problems of stability of motion in classical
  and celestial mechanics}.
\bjtitle{Russ. Math. Surv+.+}
\bvolume{18}(\bissue{6}),
\bfpage{85}
(\byear{1963})
\doiurl{10.1070/RM1963v018n06ABEH001143}
\end{barticle}
\endbibitem

\bibitem[\protect\citeauthoryear{M\"{o}ser}{1962}]{moser1962invariant}
\begin{barticle}
\bauthor{\bsnm{M\"{o}ser}, \binits{J.}}:
\batitle{On invariant curves of area-preserving mapping of an annulus}.
\bjtitle{Matematika}
\bvolume{6}(\bissue{5}),
\bfpage{51}--\blpage{68}
(\byear{1962})
\end{barticle}
\endbibitem

\bibitem[\protect\citeauthoryear{Nekhoroshev}{1977}]{Nekhoroshev:1977}
\begin{barticle}
\bauthor{\bsnm{Nekhoroshev}, \binits{N.N.}}:
\batitle{An exponential estimate of the time of stability of nearly-integrable
  hamiltonian systems}.
\bjtitle{Russ. Math. Surv+.+}
\bvolume{32}(\bissue{6}),
\bfpage{1}
(\byear{1977})
\doiurl{10.1070/RM1977v032n06ABEH003859}
\end{barticle}
\endbibitem

\bibitem[\protect\citeauthoryear{Rink}{2001}]{Rink2001}
\begin{barticle}
\bauthor{\bsnm{Rink}, \binits{B.}}:
\batitle{Symmetry and resonance in periodic {FPU} chains}.
\bjtitle{Communications in Mathematical Physics}
\bvolume{218}(\bissue{3}),
\bfpage{665}--\blpage{685}
(\byear{2001})
\end{barticle}
\endbibitem

\bibitem[\protect\citeauthoryear{Rink}{2006}]{Rink2006}
\begin{barticle}
\bauthor{\bsnm{Rink}, \binits{B.}}:
\batitle{Proof of {Nishida}'s conjecture on anharmonic lattices}.
\bjtitle{Communications in Mathematical Physics}
\bvolume{261}(\bissue{3}),
\bfpage{613}--\blpage{627}
(\byear{2006})
\doiurl{10.1007/s00220-005-1451-1}
\end{barticle}
\endbibitem

\bibitem[\protect\citeauthoryear{Henrici and
  Kappeler}{2008}]{HenriciKappeler2008Normal}
\begin{barticle}
\bauthor{\bsnm{Henrici}, \binits{A.}},
\bauthor{\bsnm{Kappeler}, \binits{T.}}:
\batitle{Results on normal forms for {FPU} chains}.
\bjtitle{Communications in Mathematical Physics}
\bvolume{278}(\bissue{1}),
\bfpage{145}--\blpage{177}
(\byear{2008})
\doiurl{10.1007/s00220-007-0387-z}
\end{barticle}
\endbibitem

\bibitem[\protect\citeauthoryear{Henrici and
  Kappeler}{2009}]{HenriciKappeler2009Resonant}
\begin{barticle}
\bauthor{\bsnm{Henrici}, \binits{A.}},
\bauthor{\bsnm{Kappeler}, \binits{T.}}:
\batitle{Resonant normal form for even periodic {FPU} chains}.
\bjtitle{Journal of the European Mathematical Society}
\bvolume{11}(\bissue{5}),
\bfpage{1025}--\blpage{1056}
(\byear{2009})
\doiurl{10.4171/JEMS/174}
\end{barticle}
\endbibitem

\bibitem[\protect\citeauthoryear{Bambusi
  et~al.}{2026}]{BambusiCaratiMaiocchi2026}
\begin{botherref}
\oauthor{\bsnm{Bambusi}, \binits{D.}},
\oauthor{\bsnm{Carati}, \binits{A.}},
\oauthor{\bsnm{Maiocchi}, \binits{A.}}:
A survey on rigorous results for the dynamics of periodic {FPU} chains.
arXiv preprint arXiv:2606.07018
(2026)
\end{botherref}
\endbibitem

\bibitem[\protect\citeauthoryear{Berchialla et~al.}{2004}]{Berchialla:2004}
\begin{barticle}
\bauthor{\bsnm{Berchialla}, \binits{L.}},
\bauthor{\bsnm{Giorgilli}, \binits{A.}},
\bauthor{\bsnm{Paleari}, \binits{S.}}:
\batitle{Exponentially long times to equipartition in the thermodynamic limit}.
\bjtitle{Physics Letters A}
\bvolume{321}(\bissue{3}),
\bfpage{167}--\blpage{172}
(\byear{2004})
\doiurl{10.1016/j.physleta.2003.11.052}
\end{barticle}
\endbibitem

\bibitem[\protect\citeauthoryear{DeLuca et~al.}{1995}]{deluca1995energy}
\begin{barticle}
\bauthor{\bsnm{DeLuca}, \binits{J.}},
\bauthor{\bsnm{Lichtenberg}, \binits{A.J.}},
\bauthor{\bsnm{Ruffo}, \binits{S.}}:
\batitle{Energy transitions and time scales to equipartition in the
  fermi-pasta-ulam oscillator chain}.
\bjtitle{Phys. Rev. E}
\bvolume{51},
\bfpage{2877}--\blpage{2885}
(\byear{1995})
\doiurl{10.1103/PhysRevE.51.2877}
\end{barticle}
\endbibitem

\bibitem[\protect\citeauthoryear{Parisi}{1997}]{Parisi:1997}
\begin{barticle}
\bauthor{\bsnm{Parisi}, \binits{G.}}:
\batitle{On the approach to equilibrium of a hamiltonian chain of anharmonic
  oscillators}.
\bjtitle{Europhysics Letters}
\bvolume{40}(\bissue{4}),
\bfpage{357}
(\byear{1997})
\doiurl{10.1209/epl/i1997-00471-9}
\end{barticle}
\endbibitem

\bibitem[\protect\citeauthoryear{De~Luca et~al.}{1999}]{Luca:1999}
\begin{barticle}
\bauthor{\bsnm{De~Luca}, \binits{J.}},
\bauthor{\bsnm{Lichtenberg}, \binits{A.J.}},
\bauthor{\bsnm{Ruffo}, \binits{S.}}:
\batitle{Finite times to equipartition in the thermodynamic limit}.
\bjtitle{Phys. Rev. E}
\bvolume{60},
\bfpage{3781}--\blpage{3786}
(\byear{1999})
\doiurl{10.1103/PhysRevE.60.3781}
\end{barticle}
\endbibitem

\bibitem[\protect\citeauthoryear{Benettin et~al.}{2013}]{Benettin:2013}
\begin{barticle}
\bauthor{\bsnm{Benettin}, \binits{G.}},
\bauthor{\bsnm{Christodoulidi}, \binits{H.}},
\bauthor{\bsnm{Ponno}, \binits{A.}}:
\batitle{The fermi-pasta-ulam problem and its underlying integrable dynamics}.
\bjtitle{Journal of Statistical Physics}
\bvolume{152}(\bissue{2}),
\bfpage{195}--\blpage{212}
(\byear{2013})
\doiurl{10.1007/s10955-013-0760-6}
\end{barticle}
\endbibitem

\bibitem[\protect\citeauthoryear{Benettin and Ponno}{2011}]{Benettin:2011}
\begin{barticle}
\bauthor{\bsnm{Benettin}, \binits{G.}},
\bauthor{\bsnm{Ponno}, \binits{A.}}:
\batitle{Time-scales to equipartition in the fermi-pasta-ulam problem:
  Finite-size effects and thermodynamic limit}.
\bjtitle{J. Stat. Phys.}
\bvolume{144}(\bissue{4}),
\bfpage{793}--\blpage{812}
(\byear{2011})
\doiurl{10.1007/s10955-011-0277-9}
\end{barticle}
\endbibitem

\bibitem[\protect\citeauthoryear{Matsuyama and
  Konishi}{2015}]{matsuyama2015multistage}
\begin{barticle}
\bauthor{\bsnm{Matsuyama}, \binits{H.J.}},
\bauthor{\bsnm{Konishi}, \binits{T.}}:
\batitle{Multistage slow relaxation in a hamiltonian system: The
  fermi-pasta-ulam model}.
\bjtitle{Phys. Rev. E}
\bvolume{92},
\bfpage{022917}
(\byear{2015})
\doiurl{10.1103/PhysRevE.92.022917}
\end{barticle}
\endbibitem

\bibitem[\protect\citeauthoryear{Spohn}{2006}]{Spohn2006Phonon}
\begin{barticle}
\bauthor{\bsnm{Spohn}, \binits{H.}}:
\batitle{The phonon boltzmann equation, properties and link to weakly
  anharmonic lattice dynamics}.
\bjtitle{Journal of Statistical Physics}
\bvolume{124}(\bissue{2--4}),
\bfpage{1041}--\blpage{1104}
(\byear{2006})
\doiurl{10.1007/s10955-005-8088-5}
\end{barticle}
\endbibitem

\bibitem[\protect\citeauthoryear{Aoki et~al.}{2006}]{AokiLukkarinenSpohn2006}
\begin{barticle}
\bauthor{\bsnm{Aoki}, \binits{K.}},
\bauthor{\bsnm{Lukkarinen}, \binits{J.}},
\bauthor{\bsnm{Spohn}, \binits{H.}}:
\batitle{Energy transport in weakly anharmonic chains}.
\bjtitle{Journal of Statistical Physics}
\bvolume{124}(\bissue{5}),
\bfpage{1105}--\blpage{1129}
(\byear{2006})
\doiurl{10.1007/s10955-006-9171-2}
\end{barticle}
\endbibitem

\bibitem[\protect\citeauthoryear{Lukkarinen}{2016}]{Lukkarinen2016}
\begin{bchapter}
\bauthor{\bsnm{Lukkarinen}, \binits{J.}}:
\bctitle{Kinetic theory of phonons in weakly anharmonic particle chains}.
In: \beditor{\bsnm{Lepri}, \binits{S.}} (ed.)
\bbtitle{Thermal Transport in Low Dimensions: From Statistical Physics to
  Nanoscale Heat Transfer}.
\bsertitle{Lecture Notes in Physics},
vol. \bseriesno{921},
pp. \bfpage{159}--\blpage{214}.
\bpublisher{Springer},
\blocation{Cham}
(\byear{2016}).
\doiurl{10.1007/978-3-319-29261-8_4}
\end{bchapter}
\endbibitem

\bibitem[\protect\citeauthoryear{Mendl et~al.}{2016}]{MendlLuLukkarinen2016}
\begin{barticle}
\bauthor{\bsnm{Mendl}, \binits{C.B.}},
\bauthor{\bsnm{Lu}, \binits{J.}},
\bauthor{\bsnm{Lukkarinen}, \binits{J.}}:
\batitle{Thermalization of oscillator chains with onsite anharmonicity and
  comparison with kinetic theory}.
\bjtitle{Physical Review E}
\bvolume{94},
\bfpage{062104}
(\byear{2016})
\doiurl{10.1103/PhysRevE.94.062104}
\end{barticle}
\endbibitem

\bibitem[\protect\citeauthoryear{Germain et~al.}{2026}]{GermainLaMenegaki2026}
\begin{botherref}
\oauthor{\bsnm{Germain}, \binits{P.}},
\oauthor{\bsnm{La}, \binits{J.}},
\oauthor{\bsnm{Menegaki}, \binits{A.}}:
A review on the kinetic theory of oscillator chains.
arXiv preprint arXiv:2606.01358
(2026)
\end{botherref}
\endbibitem

\bibitem[\protect\citeauthoryear{Onorato et~al.}{2015}]{Onorato:2015}
\begin{barticle}
\bauthor{\bsnm{Onorato}, \binits{M.}},
\bauthor{\bsnm{Vozella}, \binits{L.}},
\bauthor{\bsnm{Proment}, \binits{D.}},
\bauthor{\bsnm{Lvov}, \binits{Y.V.}}:
\batitle{Route to thermalization in the alpha-fermi-pasta-ulam system}.
\bjtitle{Proc. Natl. Acad. Sci. U.S.A.}
\bvolume{112}(\bissue{14}),
\bfpage{4208}--\blpage{4213}
(\byear{2015})
\doiurl{10.1073/pnas.1404397112}
\end{barticle}
\endbibitem

\bibitem[\protect\citeauthoryear{Lvov and Onorato}{2018}]{Lvov:2018}
\begin{barticle}
\bauthor{\bsnm{Lvov}, \binits{Y.V.}},
\bauthor{\bsnm{Onorato}, \binits{M.}}:
\batitle{Double scaling in the relaxation time in the
  $\ensuremath{\beta}$-fermi-pasta-ulam-tsingou model}.
\bjtitle{Phys. Rev. Lett.}
\bvolume{120},
\bfpage{144301}
(\byear{2018})
\doiurl{10.1103/PhysRevLett.120.144301}
\end{barticle}
\endbibitem

\bibitem[\protect\citeauthoryear{Onorato et~al.}{2023}]{Onorato:2023}
\begin{barticle}
\bauthor{\bsnm{Onorato}, \binits{M.}},
\bauthor{\bsnm{Lvov}, \binits{Y.V.}},
\bauthor{\bsnm{Dematteis}, \binits{G.}},
\bauthor{\bsnm{Chibbaro}, \binits{S.}}:
\batitle{Wave turbulence and thermalization in one-dimensional chains}.
\bjtitle{Phys. Rep.}
\bvolume{1040},
\bfpage{1}--\blpage{36}
(\byear{2023})
\doiurl{10.1016/j.physrep.2023.09.006}
\end{barticle}
\endbibitem

\bibitem[\protect\citeauthoryear{Fu et~al.}{2019}]{Fu:2019R}
\begin{barticle}
\bauthor{\bsnm{Fu}, \binits{W.}},
\bauthor{\bsnm{Zhang}, \binits{Y.}},
\bauthor{\bsnm{Zhao}, \binits{H.}}:
\batitle{Universal scaling of the thermalization time in one-dimensional
  lattices}.
\bjtitle{Phys. Rev. E}
\bvolume{100},
\bfpage{010101}
(\byear{2019})
\doiurl{10.1103/PhysRevE.100.010101}
{\href{https://arxiv.org/abs/1811.05697}{{arXiv:1811.05697}}}
{[cond-mat.stat-mech]}.
\bcomment{First posted on arXiv in 2018}
\end{barticle}
\endbibitem

\bibitem[\protect\citeauthoryear{Pistone et~al.}{2019}]{Pistone:2019}
\begin{barticle}
\bauthor{\bsnm{Pistone}, \binits{L.}},
\bauthor{\bsnm{Chibbaro}, \binits{S.}},
\bauthor{\bsnm{Bustamante}, \binits{M.D.}},
\bauthor{\bsnm{Lvov}, \binits{Y.V.}},
\bauthor{\bsnm{Onorato}, \binits{M.}}:
\batitle{Universal route to thermalization in weakly-nonlinear one-dimensional
  chains}.
\bjtitle{Math. Eng.}
\bvolume{1}(\bissue{4}),
\bfpage{672}--\blpage{698}
(\byear{2019})
\doiurl{10.3934/mine.2019.4.672}
\end{barticle}
\endbibitem

\bibitem[\protect\citeauthoryear{Fu et~al.}{2019a}]{Fu:2019}
\begin{barticle}
\bauthor{\bsnm{Fu}, \binits{W.}},
\bauthor{\bsnm{Zhang}, \binits{Y.}},
\bauthor{\bsnm{Zhao}, \binits{H.}}:
\batitle{Universal law of thermalization for one-dimensional perturbed toda
  lattices}.
\bjtitle{New J. Phys.}
\bvolume{21}(\bissue{4}),
\bfpage{043009}
(\byear{2019})
\doiurl{10.1088/1367-2630/ab115a}
\end{barticle}
\endbibitem

\bibitem[\protect\citeauthoryear{Fu et~al.}{2019b}]{PhysRevE.100.052102}
\begin{barticle}
\bauthor{\bsnm{Fu}, \binits{W.}},
\bauthor{\bsnm{Zhang}, \binits{Y.}},
\bauthor{\bsnm{Zhao}, \binits{H.}}:
\batitle{Nonintegrability and thermalization of one-dimensional diatomic
  lattices}.
\bjtitle{Phys. Rev. E}
\bvolume{100},
\bfpage{052102}
(\byear{2019})
\doiurl{10.1103/PhysRevE.100.052102}
\end{barticle}
\endbibitem

\bibitem[\protect\citeauthoryear{Wang et~al.}{2020}]{Wang:2020}
\begin{barticle}
\bauthor{\bsnm{Wang}, \binits{Z.}},
\bauthor{\bsnm{Fu}, \binits{W.}},
\bauthor{\bsnm{Zhang}, \binits{Y.}},
\bauthor{\bsnm{Zhao}, \binits{H.}}:
\batitle{Wave-turbulence origin of the instability of anderson localization
  against many-body interactions}.
\bjtitle{Phys. Rev. Lett.}
\bvolume{124},
\bfpage{186401}
(\byear{2020})
\doiurl{10.1103/PhysRevLett.124.186401}
\end{barticle}
\endbibitem

\bibitem[\protect\citeauthoryear{Sun et~al.}{2020}]{Sun:2020}
\begin{barticle}
\bauthor{\bsnm{Sun}, \binits{L.}},
\bauthor{\bsnm{Zhang}, \binits{Z.}},
\bauthor{\bsnm{Tong}, \binits{P.}}:
\batitle{Effects of weak disorder on the thermalization of
  fermi--pasta--ulam--tsingou model}.
\bjtitle{New J. Phys.}
\bvolume{22}(\bissue{7}),
\bfpage{073027}
(\byear{2020})
\doiurl{10.1088/1367-2630/ab9770}
\end{barticle}
\endbibitem

\bibitem[\protect\citeauthoryear{Wang et~al.}{2024}]{Wang:2024}
\begin{barticle}
\bauthor{\bsnm{Wang}, \binits{Z.}},
\bauthor{\bsnm{Fu}, \binits{W.}},
\bauthor{\bsnm{Zhang}, \binits{Y.}},
\bauthor{\bsnm{Zhao}, \binits{H.}}:
\batitle{Thermalization of two- and three-dimensional classical lattices}.
\bjtitle{Phys. Rev. Lett.}
\bvolume{132},
\bfpage{217102}
(\byear{2024})
\doiurl{10.1103/PhysRevLett.132.217102}
\end{barticle}
\endbibitem

\bibitem[\protect\citeauthoryear{Lin et~al.}{2025}]{Lin:2025}
\begin{barticle}
\bauthor{\bsnm{Lin}, \binits{W.}},
\bauthor{\bsnm{Fu}, \binits{W.}},
\bauthor{\bsnm{Wang}, \binits{Z.}},
\bauthor{\bsnm{Zhang}, \binits{Y.}},
\bauthor{\bsnm{Zhao}, \binits{H.}}:
\batitle{Universality classes of thermalization and energy diffusion}.
\bjtitle{Phys. Rev. E}
\bvolume{111},
\bfpage{024122}
(\byear{2025})
\doiurl{10.1103/PhysRevE.111.024122}
\end{barticle}
\endbibitem

\bibitem[\protect\citeauthoryear{Danieli et~al.}{2019}]{Danieli:2019}
\begin{barticle}
\bauthor{\bsnm{Danieli}, \binits{C.}},
\bauthor{\bsnm{Mithun}, \binits{T.}},
\bauthor{\bsnm{Kati}, \binits{Y.}},
\bauthor{\bsnm{Campbell}, \binits{D.K.}},
\bauthor{\bsnm{Flach}, \binits{S.}}:
\batitle{Dynamical glass in weakly nonintegrable {Klein--Gordon} chains}.
\bjtitle{Phys. Rev. E}
\bvolume{100}(\bissue{3}),
\bfpage{032217}
(\byear{2019})
\doiurl{10.1103/PhysRevE.100.032217}
{\href{https://arxiv.org/abs/1811.10832}{{arXiv:1811.10832}}}
\end{barticle}
\endbibitem

\bibitem[\protect\citeauthoryear{Malishava and
  Flach}{2022}]{MalishavaFlach:2022}
\begin{barticle}
\bauthor{\bsnm{Malishava}, \binits{M.}},
\bauthor{\bsnm{Flach}, \binits{S.}}:
\batitle{Lyapunov spectrum scaling for classical many-body dynamics close to
  integrability}.
\bjtitle{Phys. Rev. Lett.}
\bvolume{128}(\bissue{13}),
\bfpage{134102}
(\byear{2022})
\doiurl{10.1103/PhysRevLett.128.134102}
{\href{https://arxiv.org/abs/2109.01361}{{arXiv:2109.01361}}}
\end{barticle}
\endbibitem

\bibitem[\protect\citeauthoryear{Zhang
  et~al.}{2024}]{ZhangLandoDietzFlach:2024}
\begin{barticle}
\bauthor{\bsnm{Zhang}, \binits{W.}},
\bauthor{\bsnm{Lando}, \binits{G.M.}},
\bauthor{\bsnm{Dietz}, \binits{B.}},
\bauthor{\bsnm{Flach}, \binits{S.}}:
\batitle{Thermalization universality-class transition induced by anderson
  localization}.
\bjtitle{Phys. Rev. Research}
\bvolume{6}(\bissue{1}),
\bfpage{012064}
(\byear{2024})
\doiurl{10.1103/PhysRevResearch.6.L012064}
{\href{https://arxiv.org/abs/2308.08921}{{arXiv:2308.08921}}}
\end{barticle}
\endbibitem

\bibitem[\protect\citeauthoryear{Zwanzig}{2001}]{Zwanzig2001}
\begin{bbook}
\bauthor{\bsnm{Zwanzig}, \binits{R.}}:
\bbtitle{Nonequilibrium Statistical Mechanics}.
\bpublisher{Oxford University Press},
\blocation{New York}
(\byear{2001}).
\doiurl{10.1093/oso/9780195140187.001.0001}
\end{bbook}
\endbibitem

\bibitem[\protect\citeauthoryear{Wang et~al.}{2024}]{Wang_2024CTP}
\begin{barticle}
\bauthor{\bsnm{Wang}, \binits{Z.}},
\bauthor{\bsnm{Fu}, \binits{W.}},
\bauthor{\bsnm{Zhang}, \binits{Y.}},
\bauthor{\bsnm{Zhao}, \binits{H.}}:
\batitle{Thermalization of one-dimensional classical lattices: beyond the
  weakly interacting regime}.
\bjtitle{Commun. Theor. Phys.}
\bvolume{76}(\bissue{11}),
\bfpage{115601}
(\byear{2024})
\doiurl{10.1088/1572-9494/ad696d}
\end{barticle}
\endbibitem

\bibitem[\protect\citeauthoryear{Dudnikova et~al.}{2003}]{Dudnikova2003}
\begin{barticle}
\bauthor{\bsnm{Dudnikova}, \binits{T.V.}},
\bauthor{\bsnm{Komech}, \binits{A.I.}},
\bauthor{\bsnm{Spohn}, \binits{H.}}:
\batitle{On the convergence to statistical equilibrium for harmonic crystals}.
\bjtitle{J. Math. Phys.}
\bvolume{44}(\bissue{6}),
\bfpage{2596}
(\byear{2003})
\doiurl{10.1063/1.1571658}
\end{barticle}
\endbibitem

\bibitem[\protect\citeauthoryear{Wu et~al.}{2025}]{WuOnoratoHaniPan2025}
\begin{botherref}
\oauthor{\bsnm{Wu}, \binits{B.}},
\oauthor{\bsnm{Onorato}, \binits{M.}},
\oauthor{\bsnm{Hani}, \binits{Z.}},
\oauthor{\bsnm{Pan}, \binits{Y.}}:
Validity condition of normal form transformation for the $\beta$-{FPUT} system.
arXiv preprint arXiv:2510.04831
(2025)
\end{botherref}
\endbibitem

\bibitem[\protect\citeauthoryear{Deng and Hani}{2021}]{DengHani2021}
\begin{barticle}
\bauthor{\bsnm{Deng}, \binits{Y.}},
\bauthor{\bsnm{Hani}, \binits{Z.}}:
\batitle{On the derivation of the wave kinetic equation for {NLS}}.
\bjtitle{Forum of Mathematics, Pi}
\bvolume{9},
\bfpage{6}
(\byear{2021})
\doiurl{10.1017/fmp.2021.6}
\end{barticle}
\endbibitem

\bibitem[\protect\citeauthoryear{Deng and Hani}{2023}]{DengHani2023}
\begin{barticle}
\bauthor{\bsnm{Deng}, \binits{Y.}},
\bauthor{\bsnm{Hani}, \binits{Z.}}:
\batitle{Full derivation of the wave kinetic equation}.
\bjtitle{Inventiones Mathematicae}
\bvolume{233}(\bissue{2}),
\bfpage{543}--\blpage{724}
(\byear{2023})
\doiurl{10.1007/s00222-023-01189-2}
\end{barticle}
\endbibitem

\bibitem[\protect\citeauthoryear{Wu}{2025}]{Wu2025WKE}
\begin{botherref}
\oauthor{\bsnm{Wu}, \binits{B.}}:
Rigorous derivation of the wave kinetic equation for the $\beta$-{FPUT} system.
arXiv preprint arXiv:2506.02948
(2025)
\end{botherref}
\endbibitem

\bibitem[\protect\citeauthoryear{Nazarenko}{2011}]{2011LNP825N}
\begin{bbook}
\bauthor{\bsnm{Nazarenko}, \binits{S.}}:
\bbtitle{Wave Turbulence}.
\bsertitle{Lecture Notes in Physics},
vol. \bseriesno{825}.
\bpublisher{Springer},
\blocation{Berlin, Heidelberg}
(\byear{2011}).
\doiurl{10.1007/978-3-642-15942-8}
\end{bbook}
\endbibitem

\bibitem[\protect\citeauthoryear{Shastry and Young}{2010}]{PhysRevB.82.104306}
\begin{barticle}
\bauthor{\bsnm{Shastry}, \binits{B.S.}},
\bauthor{\bsnm{Young}, \binits{A.P.}}:
\batitle{Dynamics of energy transport in a toda ring}.
\bjtitle{Phys. Rev. B}
\bvolume{82},
\bfpage{104306}
(\byear{2010})
\doiurl{10.1103/PhysRevB.82.104306}
\end{barticle}
\endbibitem

\bibitem[\protect\citeauthoryear{Fu et~al.}{2021}]{Fu:2021}
\begin{barticle}
\bauthor{\bsnm{Fu}, \binits{W.}},
\bauthor{\bsnm{Zhang}, \binits{Y.}},
\bauthor{\bsnm{Zhao}, \binits{H.}}:
\batitle{Effect of pressure on thermalization of one-dimensional nonlinear
  chains}.
\bjtitle{Phys. Rev. E}
\bvolume{104},
\bfpage{032104}
(\byear{2021})
\doiurl{10.1103/PhysRevE.104.L032104}
\end{barticle}
\endbibitem

\bibitem[\protect\citeauthoryear{Feng et~al.}{2022}]{Feng_2022}
\begin{barticle}
\bauthor{\bsnm{Feng}, \binits{S.}},
\bauthor{\bsnm{Fu}, \binits{W.}},
\bauthor{\bsnm{Zhang}, \binits{Y.}},
\bauthor{\bsnm{Zhao}, \binits{H.}}:
\batitle{The anti-fermi--pasta--ulam--tsingou problem in one-dimensional
  diatomic lattices}.
\bjtitle{J. Stat. Mech: Theory Exp.}
\bvolume{2022}(\bissue{5}),
\bfpage{053104}
(\byear{2022})
\doiurl{10.1088/1742-5468/ac6a5a}
\end{barticle}
\endbibitem

\bibitem[\protect\citeauthoryear{Bivins et~al.}{1973}]{BIVINS197365}
\begin{barticle}
\bauthor{\bsnm{Bivins}, \binits{R.L.}},
\bauthor{\bsnm{Metropolis}, \binits{N.}},
\bauthor{\bsnm{Pasta}, \binits{J.R.}}:
\batitle{Nonlinear coupled oscillators: Modal equation approach}.
\bjtitle{J. Comput. Phys.}
\bvolume{12}(\bissue{1}),
\bfpage{65}--\blpage{87}
(\byear{1973})
\doiurl{10.1016/0021-9991(73)90169-1}
\end{barticle}
\endbibitem

\bibitem[\protect\citeauthoryear{Sholl}{1990}]{SHOLL1990253}
\begin{barticle}
\bauthor{\bsnm{Sholl}, \binits{D.}}:
\batitle{Modal coupling in one-dimensional anharmonic lattices}.
\bjtitle{Phys. Lett. A}
\bvolume{149}(\bissue{5}),
\bfpage{253}--\blpage{257}
(\byear{1990})
\doiurl{10.1016/0375-9601(90)90424-M}
\end{barticle}
\endbibitem

\bibitem[\protect\citeauthoryear{Fu et~al.}{2025}]{e27070741}
\begin{botherref}
\oauthor{\bsnm{Fu}, \binits{W.}},
\oauthor{\bsnm{Feng}, \binits{S.}},
\oauthor{\bsnm{Zhang}, \binits{Y.}},
\oauthor{\bsnm{Zhao}, \binits{H.}}:
Thermalization in asymmetric harmonic chains.
Entropy
\textbf{27}(7)
(2025)
\doiurl{10.3390/e27070741}
\end{botherref}
\endbibitem

\bibitem[\protect\citeauthoryear{Fu et~al.}{2026}]{fu2026DHPG}
\begin{barticle}
\bauthor{\bsnm{Fu}, \binits{W.}},
\bauthor{\bsnm{Wang}, \binits{Z.}},
\bauthor{\bsnm{Wang}, \binits{Y.}},
\bauthor{\bsnm{Zhang}, \binits{Y.}},
\bauthor{\bsnm{Zhao}, \binits{H.}}:
\batitle{From near-integrable to far-from-integrable: A unified picture of
  thermalization and heat transport}.
\bjtitle{Physical Review Research}
\bvolume{8}(\bissue{1}),
\bfpage{013314}
(\byear{2026})
\doiurl{10.1103/7jgy-dj3n}
\end{barticle}
\endbibitem

\bibitem[\protect\citeauthoryear{Benettin et~al.}{2023}]{Benettin2023}
\begin{barticle}
\bauthor{\bsnm{Benettin}, \binits{G.}},
\bauthor{\bsnm{Orsatti}, \binits{G.}},
\bauthor{\bsnm{Ponno}, \binits{A.}}:
\batitle{On the role of the integrable toda model in one-dimensional molecular
  dynamics}.
\bjtitle{J. Stat. Phys.}
\bvolume{190}(\bissue{8}),
\bfpage{131}
(\byear{2023})
\doiurl{10.1007/s10955-023-03147-x}
\end{barticle}
\endbibitem

\bibitem[\protect\citeauthoryear{Ooyama et~al.}{1969}]{Ooyama1969}
\begin{barticle}
\bauthor{\bsnm{Ooyama}, \binits{N.}},
\bauthor{\bsnm{Hirooka}, \binits{H.}},
\bauthor{\bsnm{Sait{\^o}}, \binits{N.}}:
\batitle{Computer studies on the approach to thermal equilibrium in coupled
  anharmonic oscillators. ii. one-dimensional case}.
\bjtitle{J. Phys. Soc. Jpn.}
\bvolume{27}(\bissue{4}),
\bfpage{815}--\blpage{824}
(\byear{1969})
\doiurl{10.1143/jpsj.27.815}
\end{barticle}
\endbibitem

\bibitem[\protect\citeauthoryear{Bocchieri and Valz-Gris}{1974}]{Bocchieri1974}
\begin{barticle}
\bauthor{\bsnm{Bocchieri}, \binits{P.}},
\bauthor{\bsnm{Valz-Gris}, \binits{F.}}:
\batitle{Ergodic properties of an anharmonic two-dimensional crystal}.
\bjtitle{Physical Review A}
\bvolume{9}(\bissue{3}),
\bfpage{1252}--\blpage{1256}
(\byear{1974})
\doiurl{10.1103/physreva.9.1252}
\end{barticle}
\endbibitem

\bibitem[\protect\citeauthoryear{Benettin et~al.}{1980}]{Benettin1980}
\begin{barticle}
\bauthor{\bsnm{Benettin}, \binits{G.}},
\bauthor{\bsnm{Vecchio}, \binits{G.L.}},
\bauthor{\bsnm{Tenenbaum}, \binits{A.}}:
\batitle{Stochastic transition in two-dimensional lennard-jones systems}.
\bjtitle{Physical Review A}
\bvolume{22}(\bissue{4}),
\bfpage{1709}--\blpage{1719}
(\byear{1980})
\doiurl{10.1103/physreva.22.1709}
\end{barticle}
\endbibitem

\bibitem[\protect\citeauthoryear{Benettin and Tenenbaum}{1983}]{Benettin1983}
\begin{barticle}
\bauthor{\bsnm{Benettin}, \binits{G.}},
\bauthor{\bsnm{Tenenbaum}, \binits{A.}}:
\batitle{Ordered and stochastic behavior in a two-dimensional lennard-jones
  system}.
\bjtitle{Physical Review A}
\bvolume{28}(\bissue{5}),
\bfpage{3020}--\blpage{3029}
(\byear{1983})
\doiurl{10.1103/physreva.28.3020}
\end{barticle}
\endbibitem

\bibitem[\protect\citeauthoryear{Benettin}{2005}]{benettin_time_2005}
\begin{barticle}
\bauthor{\bsnm{Benettin}, \binits{G.}}:
\batitle{Time scale for energy equipartition in a two-dimensional {FPU} model}.
\bjtitle{Chaos}
\bvolume{15}(\bissue{1}),
\bfpage{015108}
(\byear{2005})
\doiurl{10.1063/1.1854278}
\end{barticle}
\endbibitem

\bibitem[\protect\citeauthoryear{Benettin and Gradenigo}{2008}]{Benettin:2008}
\begin{barticle}
\bauthor{\bsnm{Benettin}, \binits{G.}},
\bauthor{\bsnm{Gradenigo}, \binits{G.}}:
\batitle{{A study of the Fermi--Pasta--Ulam problem in dimension two}}.
\bjtitle{Chaos}
\bvolume{18}(\bissue{1}),
\bfpage{013112}
(\byear{2008})
\doiurl{10.1063/1.2838458}
\end{barticle}
\endbibitem

\bibitem[\protect\citeauthoryear{Midtvedt et~al.}{2014}]{Midtvedt2014}
\begin{barticle}
\bauthor{\bsnm{Midtvedt}, \binits{D.}},
\bauthor{\bsnm{Croy}, \binits{A.}},
\bauthor{\bsnm{Isacsson}, \binits{A.}},
\bauthor{\bsnm{Qi}, \binits{Z.}},
\bauthor{\bsnm{Park}, \binits{H.S.}}:
\batitle{Fermi-pasta-ulam physics with nanomechanical graphene resonators:
  Intrinsic relaxation and thermalization from flexural mode coupling}.
\bjtitle{Phys. Rev. Lett.}
\bvolume{112}(\bissue{14}),
\bfpage{145503}
(\byear{2014})
\doiurl{10.1103/physrevlett.112.145503}
\end{barticle}
\endbibitem

\bibitem[\protect\citeauthoryear{Wang et~al.}{2018a}]{Wang2018}
\begin{barticle}
\bauthor{\bsnm{Wang}, \binits{Y.}},
\bauthor{\bsnm{Zhu}, \binits{Z.}},
\bauthor{\bsnm{Zhang}, \binits{Y.}},
\bauthor{\bsnm{Huang}, \binits{L.}}:
\batitle{Metastable states and energy flow pathway in square graphene
  resonators}.
\bjtitle{Phys. Rev. E}
\bvolume{97}(\bissue{1}),
\bfpage{012143}
(\byear{2018})
\doiurl{10.1103/physreve.97.012143}
\end{barticle}
\endbibitem

\bibitem[\protect\citeauthoryear{Wang et~al.}{2018b}]{Wang2018a}
\begin{botherref}
\oauthor{\bsnm{Wang}, \binits{Y.}},
\oauthor{\bsnm{Zhu}, \binits{Z.}},
\oauthor{\bsnm{Zhang}, \binits{Y.}},
\oauthor{\bsnm{Huang}, \binits{L.}}:
Symmetry blockade and its breakdown in energy equipartition of square graphene
  resonators.
Appl. Phys. Lett.
\textbf{112}(11)
(2018)
\doiurl{10.1063/1.5009492}
\end{botherref}
\endbibitem

\bibitem[\protect\citeauthoryear{Deng et~al.}{2026}]{Deng_2026}
\begin{barticle}
\bauthor{\bsnm{Deng}, \binits{L.}},
\bauthor{\bsnm{Fu}, \binits{W.}},
\bauthor{\bsnm{Fang}, \binits{C.}},
\bauthor{\bsnm{Wang}, \binits{Y.}},
\bauthor{\bsnm{Huang}, \binits{L.}}:
\batitle{Thermalization scaling in monolayer graphene: wave turbulence and
  beyond}.
\bjtitle{New J. Phys.}
\bvolume{28}(\bissue{2}),
\bfpage{024602}
(\byear{2026})
\doiurl{10.1088/1367-2630/ae3d5d}
\end{barticle}
\endbibitem

\bibitem[\protect\citeauthoryear{Kevrekidis and
  Cuevas-Maraver}{2019}]{Kevrekidis:2019}
\begin{bbook}
\beditor{\bsnm{Kevrekidis}, \binits{P.G.}},
\beditor{\bsnm{Cuevas-Maraver}, \binits{J.}} (eds.):
\bbtitle{A Dynamical Perspective on the $\phi^4$ Model: Past, Present and
  Future}.
\bsertitle{Nonlinear Systems and Complexity},
vol. \bseriesno{26}.
\bpublisher{Springer},
\blocation{Cham}
(\byear{2019}).
\doiurl{10.1007/978-3-030-11839-6}
\end{bbook}
\endbibitem

\bibitem[\protect\citeauthoryear{Pushkarev}{1999}]{Pushkarev:1999}
\begin{barticle}
\bauthor{\bsnm{Pushkarev}, \binits{A.}}:
\batitle{On the kolmogorov and frozen turbulence in numerical simulation of
  capillary waves}.
\bjtitle{Eur. J. Mech. B Fluids}
\bvolume{18}(\bissue{3}),
\bfpage{345}--\blpage{351}
(\byear{1999})
\doiurl{10.1016/S0997-7546(99)80032-6}
\end{barticle}
\endbibitem

\bibitem[\protect\citeauthoryear{Connaughton et~al.}{2001}]{Connaughton:2001}
\begin{barticle}
\bauthor{\bsnm{Connaughton}, \binits{C.}},
\bauthor{\bsnm{Nazarenko}, \binits{S.}},
\bauthor{\bsnm{Pushkarev}, \binits{A.}}:
\batitle{Discreteness and quasiresonances in weak turbulence of capillary
  waves}.
\bjtitle{Phys. Rev. E}
\bvolume{63},
\bfpage{046306}
(\byear{2001})
\doiurl{10.1103/PhysRevE.63.046306}
\end{barticle}
\endbibitem

\bibitem[\protect\citeauthoryear{Kartashova}{2007}]{Kartashova:2007}
\begin{barticle}
\bauthor{\bsnm{Kartashova}, \binits{E.}}:
\batitle{Exact and quasiresonances in discrete water wave turbulence}.
\bjtitle{Phys. Rev. Lett.}
\bvolume{98},
\bfpage{214502}
(\byear{2007})
\doiurl{10.1103/PhysRevLett.98.214502}
\end{barticle}
\endbibitem

\bibitem[\protect\citeauthoryear{L'vov and Nazarenko}{2010}]{Lvov:2010}
\begin{barticle}
\bauthor{\bsnm{L'vov}, \binits{V.S.}},
\bauthor{\bsnm{Nazarenko}, \binits{S.}}:
\batitle{Discrete and mesoscopic regimes of finite-size wave turbulence}.
\bjtitle{Phys. Rev. E}
\bvolume{82},
\bfpage{056322}
(\byear{2010})
\doiurl{10.1103/PhysRevE.82.056322}
\end{barticle}
\endbibitem

\bibitem[\protect\citeauthoryear{Pan and Yue}{2017}]{Pan:2017}
\begin{barticle}
\bauthor{\bsnm{Pan}, \binits{Y.}},
\bauthor{\bsnm{Yue}, \binits{D.K.}}:
\batitle{Understanding discrete capillary-wave turbulence using a
  quasi-resonant kinetic equation}.
\bjtitle{J. Fluid Mech.}
\bvolume{816},
\bfpage{1}
(\byear{2017})
\doiurl{10.1017/jfm.2017.106}
\end{barticle}
\endbibitem

\bibitem[\protect\citeauthoryear{Campbell
  et~al.}{2004}]{campbell2004localizing}
\begin{barticle}
\bauthor{\bsnm{Campbell}, \binits{D.K.}},
\bauthor{\bsnm{Flach}, \binits{S.}},
\bauthor{\bsnm{Kivshar}, \binits{Y.S.}}:
\batitle{Localizing energy through nonlinearity and discreteness}.
\bjtitle{Phys. Today}
\bvolume{57}(\bissue{1}),
\bfpage{43}--\blpage{49}
(\byear{2004})
\doiurl{10.1063/1.1650069}
\end{barticle}
\endbibitem

\bibitem[\protect\citeauthoryear{Abanin et~al.}{2019}]{RevModPhys.91.021001}
\begin{barticle}
\bauthor{\bsnm{Abanin}, \binits{D.A.}},
\bauthor{\bsnm{Altman}, \binits{E.}},
\bauthor{\bsnm{Bloch}, \binits{I.}},
\bauthor{\bsnm{Serbyn}, \binits{M.}}:
\batitle{Colloquium: Many-body localization, thermalization, and entanglement}.
\bjtitle{Rev. Mod. Phys.}
\bvolume{91},
\bfpage{021001}
(\byear{2019})
\doiurl{10.1103/RevModPhys.91.021001}
\end{barticle}
\endbibitem

\bibitem[\protect\citeauthoryear{Li and Wang}{2003}]{PhysRevLett.91.044301}
\begin{barticle}
\bauthor{\bsnm{Li}, \binits{B.}},
\bauthor{\bsnm{Wang}, \binits{J.}}:
\batitle{Anomalous heat conduction and anomalous diffusion in one-dimensional
  systems}.
\bjtitle{Phys. Rev. Lett.}
\bvolume{91},
\bfpage{044301}
(\byear{2003})
\doiurl{10.1103/PhysRevLett.91.044301}
\end{barticle}
\endbibitem

\bibitem[\protect\citeauthoryear{Lin et~al.}{2024}]{e26020156}
\begin{botherref}
\oauthor{\bsnm{Lin}, \binits{X.}},
\oauthor{\bsnm{Rondoni}, \binits{L.}},
\oauthor{\bsnm{Zhao}, \binits{H.}}:
Fluctuation relation for the dissipative flux: The role of dynamics,
  correlations and heat baths.
Entropy
\textbf{26}(2)
(2024)
\doiurl{10.3390/e26020156}
\end{botherref}
\endbibitem

\end{thebibliography}
\end{document}